\documentclass[prd,aps,twocolumn,nofootinbib,preprintnumbers,superscriptaddress,preprintnumbers,balancelastpage,longbibliography]{revtex4-2}
\usepackage[utf8]{inputenc}
\usepackage[colorlinks=true,citecolor=blue,linkcolor=blue]{hyperref}
\usepackage[normalem]{ulem}
\usepackage{amsmath,amssymb, mathrsfs,xfrac}
\usepackage{epsfig}
\usepackage{gensymb} 
\usepackage{graphicx}               
\usepackage{color}
\usepackage{slashed}
\usepackage{caption}
\usepackage{threeparttable}
\usepackage[normalem]{ulem}
\usepackage{subcaption}  
\usepackage{amsmath}
\usepackage[justification=raggedright,singlelinecheck=false]{caption}
\usepackage{enumitem} 
\usepackage{multirow}
\usepackage{placeins}
\usepackage[dvipsnames]{xcolor}
\usepackage{epstopdf}
\usepackage{soul}
\usepackage{fontawesome5}
\usepackage{tikz}
\usepackage[capitalise, english]{cleveref}
\usepackage{siunitx}
\usepackage{xspace}
\usepackage{xcolor}
\usepackage{booktabs}
\usepackage{ragged2e}
\usetikzlibrary{trees}
\usepackage{xcolor}
\definecolor{blueviolet}{rgb}{0.54, 0.17, 0.89}
\definecolor{sage}{HTML}{A9C191}   
\definecolor{green}{RGB}{0,128,0}   

\usetikzlibrary{decorations.pathmorphing}
\usetikzlibrary{decorations.markings}
\usepackage{aas_macros}
\usepackage{placeins}
\usepackage{lettrine}
\input Zallman.fd

\LettrineTextFont{\itshape}
\newcommand\myshade{80}
\colorlet{mylinkcolor}{ForestGreen}
\colorlet{mycitecolor}{ForestGreen}
\colorlet{myurlcolor}{violet}

\hypersetup{
  linkcolor  = mylinkcolor!\myshade!black,
  citecolor  = mycitecolor!\myshade!black,
  urlcolor   = myurlcolor!\myshade!black,
  colorlinks = true
}

\definecolor{jblue}{RGB}{20,50,100}
\definecolor{npurple}{RGB} {153, 51, 204}
\definecolor{wred}{RGB}{217,0,56}
\definecolor{white}{RGB}{255,255,255}

\definecolor{korange}{RGB}{235, 80,  40}
\definecolor{korange2}{RGB}{245, 100,  63}
\definecolor{kyelloworange}{RGB}{255, 210,  110}
\definecolor{kyelloworange2}{RGB}{240, 170,  90}
\definecolor{kred}{RGB}{204,  102, 153}
\definecolor{kpurple}{RGB}{153,  61, 190}
\definecolor{kpurplelight}{RGB}{213,  161, 230}

 \definecolor{tobycolour}{rgb}{.5,.0,.5}

\DeclareSIUnit\year{yr}
\DeclareSIUnit\pc{pc}
\DeclareSIUnit\ergs{ergs}
\DeclareSIUnit\msun{\ensuremath{M_\odot}}
\allowdisplaybreaks

\newcommand{\blfootnote}[1]{%
  \begingroup
  \renewcommand{\thefootnote}{}\footnote{#1}%
  \addtocounter{footnote}{-1}%
  \endgroup
}

\makeatletter
\providecommand*{\diff}%
  {\command{\lmultau}{\ensuremath{L_\mu-L_\tau}\xspace}
\new@ifnextchar^{\DIfF}{\DIfF^{}}}
\def\DIfF^#1{%
  \mathop{\mathrm{\mathstrut d}}%
    \nolimits^{#1}\gobblespace}
\def\gobblespace{%
  \futurelet\diffarg\opspace}
\def\opspace{%
  \let\DiffSpace\!%
  \ifx\diffarg(%
    \let\DiffSpace\relax
  \else
    \ifx\diffarg[%
      \let\DiffSpace\relax
    \else
        \ifx\diffarg\{%
        \let\DiffSpace\relax
      \fi\fi\fi\DiffSpace}
      
\usepackage{makecell} 
      
\usepackage{tikz,xcolor,hyperref}

\definecolor{lime}{HTML}{A6CE39}
\DeclareRobustCommand{\orcidicon}{\hspace{-1mm}
	\begin{tikzpicture}
	\draw[lime, fill=lime] (0,0) 
	circle [radius=0.16] 
	node[white] {{\fontfamily{qag}\selectfont \tiny \,ID}};
	\draw[white, fill=white] (-0.0525,0.095) 
	circle [radius=0.007];
	\end{tikzpicture}
	\hspace{-3mm}
}

\foreach \x in {A, ..., Z}{\expandafter\xdef\csname orcid\x\endcsname{\noexpand\href{https://orcid.org/\csname orcidauthor\x\endcsname}
			{\noexpand\orcidicon}}
}


\newcommand{\mytitle}{Galaxy Mergers Collectively Illuminate the $\gamma$-Ray Sky}

\begin{document}


\title{\mytitle}

\author{Jaya Doliya\orcidA{}}
\email{jayadoliya@iisc.ac.in}
\affiliation{Centre for High Energy Physics, Indian Institute of Science, C.\,V.\,Raman Avenue, Bengaluru 560012, India}

\author{Deep Jyoti Das\orcidB{}}
\email{deepjyoti@iisc.ac.in}
\affiliation{Centre for High Energy Physics, Indian Institute of Science, C.\,V.\,Raman Avenue, Bengaluru 560012, India}

\author{Subhadip Bouri\orcidC{}}
\email{subhadipb@iisc.ac.in}
\affiliation{Department of Physics, Indian Institute of Science, C.\,V.\,Raman Avenue, Bengaluru 560012, India}
\affiliation{Centre for High Energy Physics, Indian Institute of Science, C.\,V.\,Raman Avenue, Bengaluru 560012, India}

\author{Pooja~Bhattacharjee\orcidD{}}
\email{pooja.bhattacharjee@ung.si}
\affiliation{Center for Astrophysics and Cosmology, University of Nova Gorica, Vipavska 11c, SI-5270 Ajdovščina, Slovenia}

\author{Mousumi Das\orcidF{}}
\email{mousumi@iiap.res.in}
\affiliation{Indian Institute of Astrophysics, Koramangala, Bengaluru 560034, India}

\author{Ranjan Laha\orcidE{}}
\email{ranjanlaha@iisc.ac.in}
\affiliation{Centre for High Energy Physics, Indian Institute of Science, C.\,V.\,Raman Avenue, Bengaluru 560012, India}

\date{\today}


\begin{abstract}
The origin and acceleration mechanism of cosmic rays (CRs) remain fundamental open questions. Galaxy mergers are proposed as very high-energy CR accelerators, which are expected to produce high-energy (HE) $\gamma$ rays and neutrinos through interactions with the ambient gas and low-energy background radiation fields. For the first time, we systematically study the HE $\gamma$-ray emission from galaxy mergers utilising 16.7 years of \textit{Fermi} Large Area Telescope (\textit{Fermi}-LAT) data with the sample list compiled from eight survey catalogs. Our analysis finds 8 galaxy mergers that exhibit $\gamma$-ray emission with significance $\gtrsim5\sigma$ in the 1-500 GeV energy range. A stacking analysis of the remaining faint galaxy mergers yields a combined $\gamma$-ray emission detected at $\sim 35\sigma$ significance, a best-fit spectral index of $\Gamma \approx 2.07$, and an energy flux of $\sim \rm 2\times10^{-14}~erg~cm^{-2}~s^{-1}$. We compare the stacked spectral energy distributions of the galaxy mergers with the projected sensitivity of the upcoming $\gamma$-ray telescope Cherenkov Telescope Array (CTA). Furthermore, we find that 18 previously unassociated \textit{Fermi}-LAT sources are spatially coincident with galaxy mergers. Our findings establish galaxy mergers as a new class of HE $\gamma$-ray sources. Future neutrino and $\gamma$-ray observatories will be crucial to discover the particle acceleration mechanism in these newly identified CR sources. 
\end{abstract}

\maketitle
\blfootnote{The first three authors contributed equally to this work.}
\section{Introduction}
\label{sec:Introduction}
Cosmic rays (CRs) are the most energetic charged nuclei ever detected on Earth, with the highest recorded energy being $\sim3.2 \times 10^{20}$\,eV~\cite{Linsley:1963km, PierreAuger:2020kuy, TelescopeArray:2023sbd}. Ultra-high-energy (UHE) CRs have ignited significant interest in the search for their production sites and acceleration mechanisms~\cite{2012APh....39...52D, Blasi:2013rva, Baring:2004qc,2012SSRv..173..491S, Ostrowski:2001ej, Parker:1958zza, Aloisio:2022xzy, Zweibel:2013qsz, Sommers:2008ji}. As CRs are charged particles~\cite{PhysRev.43.87}, their trajectories are deflected by galactic and intergalactic magnetic fields, causing them to arrive at Earth nearly isotropically~\cite{PhysRev.47.817,1966ApJ...146..480J, Adriani:2015kfa, LaVacca:2016tqq, KASCADEGrande:2017vwf, Fermi-LAT:2017vjf, Fermi-LAT:2019xgo, HAWC:2018wju, PierreAuger:2023fcr, IceCube:2023gpr, IceCube:2024pnx, TelescopeArray:2024oux}. Thus, it is extremely challenging to trace their origin\,\cite{2011ARA&A..49..119K, Luque:2020oek,deMelloNeto:2020xfa, Lipari:2003es, Aloisio:2022xzy, Evoli:2023kxd}.

CRs can interact with matter and radiation in the vicinity of their sources, as well as with low-energy radiation backgrounds, such as the cosmic microwave background (CMB) and the extragalactic background light (EBL), while travelling through intergalactic space, leading to the production of $\gamma$-rays and neutrinos~\cite{Achterberg:1999vr, Aps1990, 2013A&A...553A..34D, Hussain:2023jgt}. Gamma-rays and neutrinos are both electrically neutral particles; therefore, they are undeflected by the magnetic fields of astrophysical sources, host galaxies, the intergalactic medium, and the Milky Way on their way to Earth. However, very high-energy (VHE) $\gamma$ rays ($\gtrsim$ 10 TeV) can be attenuated during propagation over cosmological distances due to interactions with background photons through electron-positron pair production. For instance, at 1 TeV and 1 PeV, where EBL and CMB absorptions dominate, the mean free paths of photons are $\sim$ 370 Mpc and $\sim$ 10 kpc, respectively. In contrast, photons with energies of $\mathcal{O}(100 \,\rm GeV)$ have a mean free path of 0.1 -- 1\,Gpc, allowing them to reach the Earth largely unattenuated. These electron-positron pairs can up-scatter the ambient low-energy diffuse photons via inverse Compton scattering, initiating an electromagnetic cascade. This process partially compensates for the attenuation, and leads to the production of comparatively low-energy $\gamma$-rays of $\mathcal{O}$(10 -- 100 GeV) that can reach the Earth~\cite{Berezinsky:1975zz, Fiorillo:2025kuh, Gould:1967zza, Stecker:2005qs,2010A&A...515A..19K, Dominguez:2010bv, Dwek:2012nb, Stecker:2016fsg, Franceschini:2017iwq}. Gamma-rays can provide a plethora of information about CR production factories,  underlying production mechanisms, the nature of extreme astrophysical environments, and potential new fundamental physics~\cite{Funk:2015ena, DeAngelis:2018lra, Stecker:2025lyd, Funk:2015ena, Prandini:2022wcb, Pueschel:2021yjw}.

A wide range of astrophysical sources have been explored in the search for $\gamma$-rays. Galactic astrophysical objects from which high-energy (HE) $\gamma$-rays have either been detected or are expected to be emitted include supernova remnants~\cite{10.1093/mnras/182.3.443,2013A&A...553A..34D, Giuliani:2024fpq}, stellar wind binaries~\cite{2011ApJ...743....7Z, VERITAS:2021ogb, HESS:2006csb, Aharonian:2007yb, MAGIC:2008eqb, Marcote:2018rsy}, star-cluster winds~\cite{Bykov:2014asa, Peron:2024bok, Blasi:2025yjl, Blasi:2023quw, Pandey:2024cib,2013ICRC...33.1216K, Peron:2024lfo, Maurin:2016hhl}, microquasars~\cite{Bosch-Ramon:2004jlg, Bosch-Ramon:2006tax, Alfaro:2024cjd, F.Aharonian_2006, Ohira:2024qtr,2002A&A...390..751H, Neronov:2024ycp, Zhang:2025tew}, binary neutron star mergers~\cite{LIGOScientific:2017ync, RODRIGUES201910, Guo:2025wkt, Farrar:2024zsm}, pulsars~\cite{Blasi:2000xm, Lopez-Coto:2022igd}, pulsar wind nebulae (PWNe)~\cite{Xiao:2024cgh, Bednarek:2003tk}, magnetars~\cite{Arons:2002yj, Sousa:2024mbz, Kotera:2011vs}, and many others. Extragalactic confirmed sources and potential candidates include active galactic nuclei (AGN)~\cite{Rieger:2022qhs,deOliveira:2024juz, HESS:2012xzq}, starburst galaxies~\cite{Anchordoqui:1999cu, Romero:2018mnb, VERITAS:2025jtu,2009Sci...326.1080A, Peretti:2018tmo, Thompson:2006qd,2021AAS...23735101R, HESS:2005zyi}, blazars~\cite{Tavecchio:2016dcj, Sasse:2025anm, Murase_2012, Das:2021cdf}, gamma-ray bursts (GRBs)~\cite{Milgrom:1995um, DERUJULA2025139641}, fast radio bursts (FRBs)~\cite{Principe:2023bow}, fast black hole winds~\cite{Fermi-LAT:2021ibj, Karwin:2021kds}, quasars~\cite{1985ICRC....1..285L, Principe:2021muz, Principe:2021xeo, Moncada:2017hvq}, hot coronae in radio-quiet AGN~\cite{Fermi-LAT:2025chw, Inoue:2021tcn}, and many others.

We investigate galaxy mergers as HE $\gamma$-ray emitters using 16.7 years of \textit{Fermi} Large Area Telescope (\textit{Fermi}-LAT) data. Theoretically, these are well-motivated candidates for producing HE $\gamma$-ray emission~\cite{2011arXiv1104.0956V, Kashiyama:2014rza, Yuan:2017dle, Yuan:2018erh}. In Refs.\,\cite{Kashiyama:2014rza, Yuan:2017dle}, the authors show that massive galaxy mergers are capable of accelerating cosmic particles up to $\sim$ EeV energies, corresponding to the second knee of the CR spectrum with a spectral index $\Gamma$ $\approx$ 2.
Shock-accelerated CRs can interact with ambient matter and low-energy photons. These interactions lead to the production of pions ($\pi^{\pm,0}$) via hadronuclear $p + p/A \rightarrow \mathrm{n}[\pi^+ + \pi^- + \pi^0] + X$ and photohadronic processes $p + \gamma \rightarrow \Delta^+ \rightarrow \pi^0/\pi^+ + p/n$. Here, $p/\,A,\,n$, and $X$ denote the proton/target nuclei, neutron, and final-state particles, respectively. The multiplicity of the produced $\pi^+, \pi^-$, and $\pi^0$ is represented by $\mathrm{n}$. The charged pions predominantly decay into muons and neutrinos: $\pi^{\pm} \rightarrow \mu^{\pm} + \nu_{\mu}/\bar{\nu}_{\mu}$, and neutral pions decay into photons: $\pi^0 \rightarrow \gamma\gamma$. CR electrons and positrons can also produce HE $\gamma$-rays through inverse Compton scattering off the low-energy radiation background. It has been shown that collisions between galaxy groups and clusters can generate large-scale shocks that can accelerate CR nuclei beyond the PeV range, producing hard spectra with indices in the range $1.5 \lesssim \Gamma \lesssim 2.0$. The IceCube Neutrino Observatory has detected a diffuse astrophysical neutrino flux from all directions in the sky~\cite{IceCube:2013cdw, IceCube:2013low, IceCube:2014stg, IceCube:2015gsk, IceCube:2015qii, Stettner:2019tok, Stachurska:2019wfb, IceCube:2020acn, Abbasi:2021qfz, IceCube:2020wum, IceCube:2021rpz, IceCube:2024fxo, IceCube:2024nhk, IceCube:2025zyb}. These studies\,\cite{Kashiyama:2014rza, Yuan:2017dle} suggest that galaxy mergers could account for a significant fraction of the astrophysical diffuse neutrino flux detected by IceCube. Therefore, observations of HE neutrinos and $\gamma$-rays in the direction of galaxy mergers could be a signature of CR acceleration in these systems. 

A comprehensive search for HE neutrinos from galaxy mergers has been conducted in Ref.\,\cite{Bouri:2024ctc} utilizing IceCube data. This study found no statistically significant correlation between the studied galaxy mergers and IceCube-detected neutrinos. As there is a tight correlation among CRs, neutrinos, and $\gamma$-rays, it is essential to further search for the $\gamma$-rays from galaxy mergers to examine the HE astrophysical properties of this class of sources.  
 
For the first time to our knowledge, we conducted a systematic stacking analysis of $\gamma$-ray emission from galaxy mergers utilizing eight catalogs. We find 8 galaxy mergers as significant $\gamma$-ray emitters at $\gtrsim5\sigma$ confidence, that are not included in the \textit{Fermi}-LAT 4FGL-DR4 catalog\,\cite{Ballet:2023qzs}. Our stacking analysis with faint target sources demonstrates that, as a population, galaxy mergers collectively exhibit a $\gamma$-ray signal at the level of $\sim35\sigma$. Furthermore, by examining the spatial overlap between galaxy merger candidates and \textit{Fermi}-LAT unassociated sources, we find positional coincidences for 18 unassociated $\gamma$-ray sources -- with at least one galaxy merger from our sample list within the 95\% angular uncertainty region of the corresponding \textit{Fermi}-LAT source. 

The paper is structured as follows. In section \ref{sec:gal_catalog}, we discuss the galaxy merger catalogs considered and the source selection used in this study. In section \ref{sec:lat_data}, we provide an overview of the \textit{Fermi}-LAT data analysis formalism. In section \ref{sec:results}, we present our results from individual and stacking analyses and discuss the association between galaxy mergers and \textit{Fermi}-LAT unassociated sources. Finally, in section \ref{sec:discussion}, we describe the implications of our findings and conclude with a future outlook.

\section{Galaxy merger catalogs}
\label{sec:gal_catalog}
Galaxy mergers are dramatic cosmic events in which galaxies that are pulled together by gravity interact, and eventually fuse into a new, larger galaxy. Over cosmological timescales, a significant number of massive galaxies undergo at least one major merger -- typically involving two galaxies of comparable masses -- as well as several minor mergers~\cite{Conselice:2014joa}. In a galaxy merger, the vast distances between stars within a galaxy make direct stellar collisions incredibly rare. Instead, because of the enormous gravitational fields, gas and dust interact. Such mergers of gaseous structures can generate shock waves capable of accelerating CRs up to $\sim$ $\mathcal{O}$(EeV) energies\,\cite{Kashiyama:2014rza, Yuan:2017dle}.

Galaxy mergers lead to several detectable signatures across various observational wavelengths. A primary indicator is asymmetry in galaxy shape and morphology. The maximum asymmetry observed is typically seen in ongoing mergers, compared to the early and late stages of interaction. Powerful gravitational interactions can pull out long streams of stars and gas, producing strong tidal tails and bridge-like structures~\cite{Toomre:1972vt, Conselice:2014joa}. The intense compression of gas during a merger often triggers starbursts~\cite{2023MNRAS.518.3261P}. Additionally, the funnelling of gas and dust toward the galactic core during a merger can fuel supermassive black hole growth and AGN activity~\cite{Ellison:2013fla}. The presence of dual AGNs can also be a strong indicator of a merger~\cite{2004AJ....128...62G}. In Ref.\,\cite{Patton:2005qd}, the authors found evidence for a correlation between asymmetry and star forming properties in galaxy pairs. Recently, the James Webb Space Telescope (JWST) has revealed a large population of galaxy mergers that had not been detected previously\,\cite{2025MNRAS.540..774D,2024arXiv241104944D,2024ApJ...976L...8R,rodighiero2024optically,2024MNRAS.533.4472D}.

\begin{figure*}[htbp]
    \centering
    \includegraphics[width=0.9\linewidth]{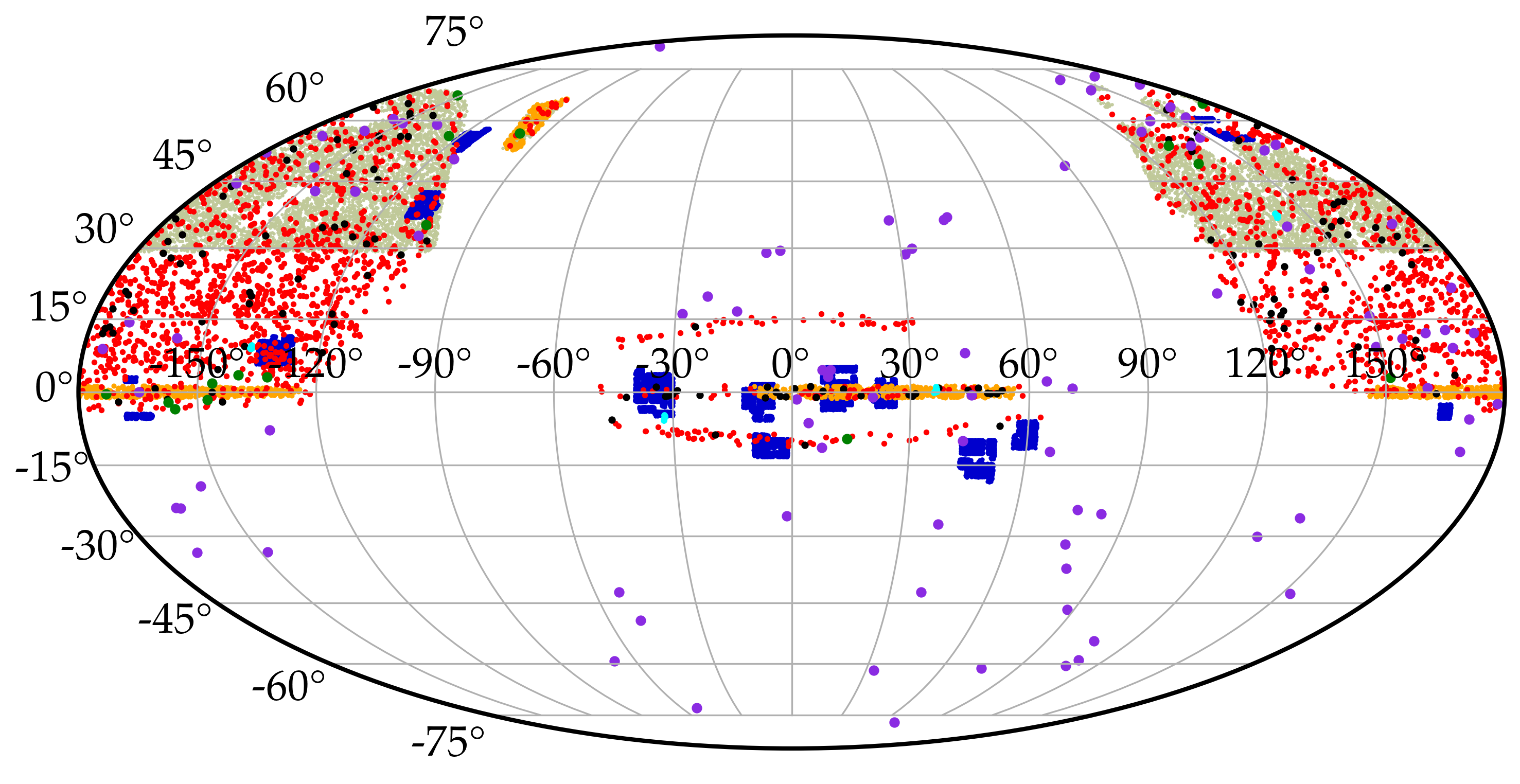}
    \caption{Locations of 31,464 galaxy mergers considered in the analysis from eight different catalogs are shown in the equatorial coordinate system, where the horizontal axis and the vertical axis represent the Right Ascension (RA) and Declination (Dec), respectively. The sage, blue, cyan, purple, black, orange, red and green colored points represent the galaxy mergers from UNIONS DR5-SDSS DR7\,\cite{2024MNRAS.533.2547F}, \href{https://cdsarc.cds.unistra.fr/viz-bin/cat/J/ApJS/181/233}{J/ApJS/181/233}\,\cite{2009yCat..21810233H}, \href{https://cdsarc.cds.unistra.fr/viz-bin/cat/J/AJ/130/2043}{J/AJ/130/2043}\,\cite{Yee:2000qj}, \href{https://cdsarc.u-strasbg.fr/viz-bin/Cat?J/AJ/128/62}{J/AJ/128/62}\,\cite{2004AJ....128...62G}, Dual AGN\,\cite{2023MNRAS.524.4482B}, \href{https://cdsarc.cds.unistra.fr/viz-bin/cat/J/AJ/127/1883}{J/AJ/127/1883}\,\cite{Allam:2003me}, \href{https://cdsarc.cds.unistra.fr/viz-bin/cat/J/MNRAS/479/415}{J/MNRAS/479/415}\,\cite{2018MNRAS.479..415A} and \href{https://cdsarc.cds.unistra.fr/viz-bin/cat/J/MNRAS/388/1537}{J/MNRAS/388/1537}\,\cite{McIntosh:2007hh} catalogs, respectively.}
    \label{fig:Merger catalog}
\end{figure*}

We use galaxy merger information from eight catalogs:
\begin{enumerate}[label=(\roman*), labelwidth=1.5em, labelsep=0.5em, leftmargin=!, align=left, font=\itshape, itemsep=1em]
    \item We include the catalog of 13,448 ``high-confidence galaxy merger candidates" reported in Ref.\,\cite{2024MNRAS.533.2547F}. The authors introduced \textsc{Mummi}, a hybrid deep-learning framework designed to identify galaxy mergers from images. It combines Convolutional Neural Networks (CNN) and Vision Transformer (ViT) deep learning algorithms, that were trained on realistic simulated data from IllustrisTNG100-1, and applied to real galaxy survey data from the overlapping regions of  UNIONS DR5 and SDSS DR7. Throughout this paper, sources from this catalog are labeled with the prefix “UN". These galaxy mergers span a redshift range  $z \approx 0.01-0.3$. Candidates from this catalog are represented by sage-coloured dots (\textcolor{sage}{$\bullet$}) in Fig.\,\ref{fig:Merger catalog}. 
    \item In Ref.\,\cite{2009yCat..21810233H}, Hwang and Chang identified 15,147 galaxy mergers by employing a morphological pattern recognition algorithm on 422 square degrees of the sky using data from the Red Sequence Cluster Survey 2 (RCS2)\,\cite{Yee:2007if}. This survey, conducted by the Canada–France–Hawaii Telescope (CFHT), targets galaxy clusters reaching redshifts of $z \sim 1$. Based on structural features of the identified sources, they classified thse into three categories: mergers (M), close pairs (CP), and close multiples (CM). For our analysis, we include a subset of sources labeled as mergers (M), resulting in 13,971 galaxy mergers from this catalog (\href{https://cdsarc.cds.unistra.fr/viz-bin/cat/J/ApJS/181/233}{J/ApJS/181/233}) being incorporated into our sample list. These mergers are represented by blue dots (\textcolor{blue}{$\bullet$}) in Fig.\,\ref{fig:Merger catalog}.
    \item We also include 35 dynamically close galaxy pairs investigated in Ref.\,\cite{Patton:2005qd}, which found that approximately half of these galaxy pairs are in their merging phase. This sample (\href{https://cdsarc.cds.unistra.fr/viz-bin/cat/J/AJ/130/2043}{J/AJ/130/2043}) was drawn from the Canadian Network for Observational Cosmology 2 (CNOC2) field galaxy redshift survey~\cite{Yee:2000qj}, conducted with the CFHT. These pairs are in the redshift range $0.1 \lesssim z \lesssim 0.6$. In Fig.\,\ref{fig:Merger catalog}, these pairs are represented by cyan dots (\textcolor{cyan}{$\bullet$}).
    \item We use the catalog of Double Nucleus Disk Galaxies (\href{https://cdsarc.u-strasbg.fr/viz-bin/Cat?J/AJ/128/62}{J/AJ/128/62}), which provides a list of 107 candidates identified as minor mergers, as detailed in Ref.\,\cite{2004AJ....128...62G}. The criteria for selecting these galaxies include $z\lesssim 0.05$ or absolute magnitude $< 18$. Only disk-like galaxies without strong tidal distortion were included to avoid complications from major mergers and to focus on minor mergers. These mergers are represented by purple dots (\textcolor{blueviolet}{$\bullet$}) in Fig.\,\ref{fig:Merger catalog}. 
    \item We utilise 159 confirmed dual-nucleus galaxies (DNGs), including three systems hosting triple-nucleus systems from Ref.\,\cite{2023MNRAS.524.4482B}. In this study, the authors employ the Graph-Boosted Iterated Hill Climbing (GOTHIC) algorithm to detect double- and multi-nucleus galaxies within a sample of 1 million SDSS DR16 galaxies. We label objects from this catalog with the prefix ``GT''. The redshift range of DNGs in this catalog is $0.023 \lesssim z \lesssim 0.260$. In Fig.\,\ref{fig:Merger catalog}, these sources are shown by black dots (\textcolor{black}{$\bullet$}). 
    \item We use 1,479 pairs of galaxy mergers from the catalog (\href{https://cdsarc.cds.unistra.fr/viz-bin/cat/J/AJ/127/1883}{J/AJ/127/1883}), identified through an automated routine search over 462 square degrees of the Sloan Digital Sky Survey Early Data Release (SDSS EDR), detailed in Ref.\,\cite{Allam:2003me}. The redshift range of the survey extends up to $z \sim 1$. The considered sources are shown as orange dots (\textcolor{orange}{$\bullet$}) in Fig.\,\ref{fig:Merger catalog}. 
    \item We select a sample of galaxy mergers from the catalog (\href{https://cdsarc.cds.unistra.fr/viz-bin/cat/J/MNRAS/479/415}{J/MNRAS/479/415}) presented in Ref.\,\cite{2018MNRAS.479..415A}, where a deep CNN algorithm was trained for automatic visual identification of galaxy mergers using SDSS DR7 images of galaxies selected and labeled based on GZ1~\cite{Darg:2009rc}. The authors used a 4-fold cross-validation method, resulting in four independent classifier outputs $p_m \in (0,1)$, where $p_m = 0$ means a non-interacting system and $p_m = 1$ is a merger system. We select only those galaxies for which all four classifier outputs yielded $p_m \approx 1$. This criterion results in a set of 2,244 galaxy mergers in $0.01 \lesssim z \lesssim 1.3469$. We label sources from this catalog with the prefix ``GZ''. These galaxy mergers are shown as red dots (\textcolor{red}{$\bullet$}) in Fig.\,\ref{fig:Merger catalog}.
    \item We use 21 progenitors of massive central merger systems found in dense environments from the catalog (\href{https://cdsarc.cds.unistra.fr/viz-bin/cat/J/MNRAS/388/1537}{J/MNRAS/388/1537})~\cite{McIntosh:2007hh}. Merger candidates are identified on the basis of structural distortions such as broad and diffuse tidal tails, asymmetric structures associated with both progenitors and with a projected separation of $\lesssim 30\,\rm kpc$. The redshift range of these progenitors lies within $0.027 \lesssim z \lesssim 0.12$. These sources are represented by green dots (\textcolor{green}{$\bullet$}) in Fig.\,\ref{fig:Merger catalog}. 
    
\end{enumerate}  

A total of 31,464 galaxy mergers were identified from these aforementioned catalogs. 

A new catalog of quasar pairs, constructed from the DESI DR1 quasar sample, was recently published~\cite{Jing:2025txo}, identifying 1,842 candidate pairs at redshifts $z\sim1–2.5$. These systems show a nearly flat projected-separation distribution beyond 20\,kpc, indicating that they are relatively stable systems and associated, but not yet dynamical coalescing. However, a thorough analysis of these quasar pairs is necessary before any definitive conclusions can be drawn.  We did not include these in our analysis.

\subsection*{Additional cuts for final source selection}
\label{sec:sample cut}
Our analysis focuses on $\gamma$-ray photons in the energy range of 1--500 GeV. The point spread function (PSF) of the \textit{Fermi}-LAT at the lowest end of our energy range is $\sim0.8\degree$ for front events and $\sim1\degree$ for back events at the $68\%$ containment level~\cite{psf1}. Therefore, to minimize the contamination from known $\gamma$-ray sources in the 4FGL-DR4 catalog, we choose only those galaxy mergers that are 1\degree\,and 5\degree\,away from 4FGL-DR4 point and extended sources, respectively. Numerous galaxy mergers have smaller angular separations from each other compared to the PSF in the energy range we are interested in, which hinders the individual resolution by the \textit{Fermi}-LAT. These will lead to overestimation in the stacked signal, see subsection\,\ref{sec:stacking analysis}. To address this, we pick only one galaxy merger within each 0.8\degree \,region.

Application of these separation criteria to the entire catalog reduces our sample to 1,875 galaxy mergers. We conduct our analysis on this refined set of galaxy mergers\footnote{The final list of 1,875 galaxy mergers analyzed is available at this \faGithub\ \href{https://github.com/Jaya-Doliya/Gamma-Ray-Search-From-Galaxy-Mergers-using-Fermi-LAT}{Github repository}}

\section{\textit{Fermi}-LAT data analysis}
\label{sec:lat_data}

To investigate $\gamma$-ray signals from galaxy mergers, we utilize data from the \textit{Fermi}-LAT, a key instrument on board the Fermi Gamma-ray Space Telescope, which has been operating for 17\,years in low Earth orbit at an altitude of 535\,km. The telescope’s tracker, composed of tungsten and silicon strips, enables the conversion of incoming $\gamma$-ray photons into electron-positron pairs via pair production  (\(\gamma + A \rightarrow  e^- + e^+ + A\)) and records their trajectories. The resulting electrons and positrons deposit their energy in the calorimeter. By reconstructing their paths and measuring the deposited energies, the instrument determines the direction and energy of the incoming $\gamma$-ray photon~\cite{2009ApJ...697.1071A, Fermi-LAT:2012fsm}.

\subsection{Pre-processing: Single Merger Analysis}
\label{sec:Single-source-analysis}

We employ a two-step analysis pipeline. First, we perform a standard binned likelihood analysis for each region of interest (ROI) individually to search for $\gamma$-ray emission associated with galaxy mergers. In the second step, we conduct a stacking analysis on the remaining faint sources (with detection significances $< 5\sigma$) to investigate potential cumulative emission of $\gamma$ rays from galaxy mergers in subsection~\ref{sec:stacking analysis}. This method of stacking has been successfully used for a variety of sources~\cite{Fermi-LAT:2011vow, Paliya:2019anr, Paliya:2021sbv, Khatiya:2023lkg, Fermi-LAT:2018lqt, Ajello:2020zna, Fermi-LAT:2021ibj, McDaniel:2023vsh}.

We use all-sky weekly data~\cite{allskydata} collected by the \textit{Fermi}-LAT over a span of 16.7 years, from August 4, 2008, to February 21, 2025 (in MET 239557417s--761845417s), to analyze target galaxy mergers in the 1--500 GeV energy range. The analysis is performed using \texttt{Fermipy} v1.3.1, built on \texttt{Fermitools} v2.2.0. Events are binned into eight logarithmic bins per energy decade, with a spatial bin size of 0.08\degree. The ROI is a 10\degree\,$\times$\,10\degree\ square, centered on the coordinates of each target galaxy merger. A 90\degree\ zenith angle cut is applied to exclude $\gamma$ rays originating from the Earth's limb~\cite{Fermi-LAT:2009dsu}. We use event class \,\texttt{P8R3\_SOURCE} (128) and event type \texttt{FRONT/BACK} (3), together with the corresponding instrument response function \texttt{P8R3\_SOURCE\_V3}~\cite{Atwood:2013dra}. We use the standard \texttt{DATA\_QUAL\ >\ 0\ \&\&\ LAT\_CONFIG==1} to select the good time intervals (GTIs) from the data for analysis. The energy dispersion correction (\texttt{edisp\_bins = -1}) is enabled for all sources except the isotropic component~\cite{edspdins}. A full-sky livetime cube was generated for the selected time range and energy range, and consistently applied to all ROIs.
We model the astrophysical $\gamma$-ray background using the fourth \textit{Fermi}-LAT source catalog (4FGL-DR4)~\cite{Ballet:2023qzs}, which includes $\gamma$-ray sources detected over 14 years of observations. Specifically, we use the most recent version of the source list, \texttt{gll\_psc\_v35}, corresponding to 4FGL-DR4. The model incorporates the Galactic diffuse emission model \texttt{(gll\_iem\_v07)} along with its corresponding extragalactic isotropic diffuse emission model \texttt{(iso\_P8R3\_SOURCE\_V3\_v1)}. Additionally, to mitigate the effects of photon leakage from sources outside the ROI due to the PSF of the detector, the model includes all 4FGL-DR4 sources within the extended 15\degree\,$\times$\,15\degree\, region. During the model fitting, the normalisation and spectral index of the Galactic component and the normalisation of the isotropic component are set free. We also allow the normalisation parameters for sources with $\rm TS > 25$ within 5\degree\ and $\rm TS > 500$ within 7\degree\ of the ROI from the centre to vary freely. The target galaxy mergers are modeled as a point source with a \texttt{PowerLaw}~\cite{{Fermi_SourceModelDefinitions2}} energy spectrum $dN/dE=N_{0}\,(E/E_{0})^{-\Gamma}$, where $N_{0}$ is the normalization of photon flux, $E_{0}$ is an arbitrary pivot energy scale, and $\Gamma$ is the spectral index. Both the normalisation and the spectral index are allowed to vary.

We adopt the maximum likelihood ratio method to evaluate the significance of a source. The Test Statistic (TS) is used to quantify the fit when a source is added to the model. It is defined as:\begin{equation} \label{eq:TS}
   {\rm TS} =  2 \ln \Big[\frac{L_{s,max}}{L_{0,max}}\Big] \,,
\end{equation}
where $L_{0,max} $ is the maximum likelihood achieved for a model in the absence of the new source (the null hypothesis), and $L_{s,max}$ is the maximum likelihood obtained in the presence of the same source in the model. According to Wilk's theorem~\cite{Wilks:1938dza}, in the large sample size limit, the TS asymptotically follows a $\chi^2$ distribution under the null hypothesis. Therefore, the $\sqrt{\rm TS}$ value provides an approximate measure of the detection significance ($\sigma$) for the new source. 

Since the time span of the data extends beyond that of the 4FGL-DR4 catalog (released on August 2, 2022), we use \texttt{gta.find\_source} to search for the new $\gamma$-ray sources in the ROI. It generates the TS map of the ROI with a test model of spectral index 2.0. When an excess emission with $\rm TS > 25$ is detected, it is added to the model with the \texttt{PowerLaw} spectrum. Then a second TS map is generated with the updated model. This process is repeated until all significant excesses above the background are identified and incorporated into the model. If no source is found in the vicinity of the center of ROI, we add a source at the center of ROI with a \texttt{PowerLaw} model, and localize it using \texttt{gta.localize}. Additionally, if any source has a high Test Statistic (TS $>$ 15) with an unphysically steep spectral index ($\Gamma>$  3.0), the source is removed from the model, and a TS map is generated with a test model of spectral index 2.8 to search for new sources. After additional sources are re-identified, the main source is re-added to the existing source model and refitted to get physically reasonable results. Finally, a full likelihood fit is performed to optimize all free parameters within the ROI and to derive the spectral parameters of the source of interest.

\subsection{Stacking Analysis}
\label{sec:stacking analysis}
In the second phase of the analysis, we construct a two-dimensional TS profile for faint target galaxy mergers by varying the integrated photon flux and the spectral index as free parameters over a range. Assuming that the entire population of galaxy mergers can be adequately characterized by an average spectral index and integrated flux value using a power-law model (\texttt{PowerLaw2}). The integrated photon flux is sampled in 50 logarithmically spaced bins ranging from $10^{-12}$ to $10^{-10}$ ph cm$^{-2}$ s$^{-1}$. The lower edge of this range lies below the \textit{Fermi}-LAT sensitivity threshold ($10^{-11}$ ph cm$^{-2}$ s$^{-1}$ at 1 GeV), ensuring sufficiently low flux that represents the null hypothesis in TS calculations in Eq.\,\ref{eq:TS}. The spectral index is varied from 1.5 to 3.0 in increments of 0.03.

 We employ the \texttt{PowerLaw2}\,\cite{FermiSourceModels2018} spectral model for the stacking analysis, defined as \\
\begin{equation}
    \frac{dN}{dE} = \frac{N(1-\Gamma)E^{-\Gamma}}{E^{1-\Gamma}_{max} - E^{1-\Gamma}_{min}} \,.
\end{equation}
Here, $N$ is the integrated photon flux and $\Gamma$ is the spectral index. $E_{min}$ and $E_{max}$ are fixed at the lowest and highest energies of our interest range.  \texttt{PowerLaw2} directly parametrizes normalization $N$ over the energy range $E_{min}$ to $E_{max}$, so it can directly propagate error to $N$. We generate a two-dimensional TS profile for each faint merger in the source list. Finally, taking advantage of the additive nature of the log-likelihood, we stack all individual TS profiles and add the corresponding TS values to evaluate whether the faint sample produces a statistically significant signal above the background.

An alternative approach for stacking $\gamma$-ray sources involves combining their photon count maps. In this method, the count maps from each ROI are aligned and summed to produce a stacked count map. This combined map can then be analyzed using likelihood techniques to estimate the overall signal significance. This stacking method has been used in~\cite{Sambruna:2007qzx,Torres:2007pua,Cillis:2006xn,Anderson:2015dpc,Song:2024bnw,Huber:2013cia}.

\subsection{Background Stacking}
\label{sec:background stacking}
To ensure the robustness of our findings and the accuracy of the background modeling, we apply the same two step pipeline to blank-sky regions:\,\,areas of the sky that are distant from known and potentially unresolved $\gamma$-ray sources. We choose regions which are at high galactic latitude $(|b|>30\degree)$, and are located at least $1\degree$ and $5\degree$ away from point and extended 4FGL-DR4 sources, respectively. We further exclude regions lying within 0.5° of sources listed in BZCat~\cite{Massaro:2008ye}, CRATES~\cite{Healey:2007by}, and WIBRaLS~\cite{DAbrusco:2014nij} catalogs to minimize the contribution from unresolved $\gamma$-ray sources to the background model~\cite{Carlson:2014nra,McDaniel:2023bju}. To prevent overlap with our galaxy merger sample list, we also exclude regions lying within $1\degree$ of our combined catalog. After applying these cuts, we perform the same analysis formalism on 3,134 blank-sky regions. 

\section{Results}
\label{sec:results}

We present the outcomes of our analysis. We show the TS distribution of the considered sources and the TS maps of the significant sources that we find in our analysis. Also, we present the light curve for the most significant source, RCS2338 14991. Next, we discuss the findings of some theoretically motivated galaxy mergers, which were mentioned in previous studies~\cite{Kashiyama:2014rza, Yuan:2017dle, Yuan:2018erh}. In the stacking analysis, we discuss the best-fit photon flux and spectral index of the considered population of sources, compare them with randomly selected blank sky positions, and discuss the implications. Additionally, we also show the stacked spectral energy distribution (SED) of the sources.

\subsection{Individual Galaxy Merger Analysis}
\label{sec:single-source-results}

\begin{figure}[!htbp]
    \centering
    \includegraphics[width=1\linewidth]{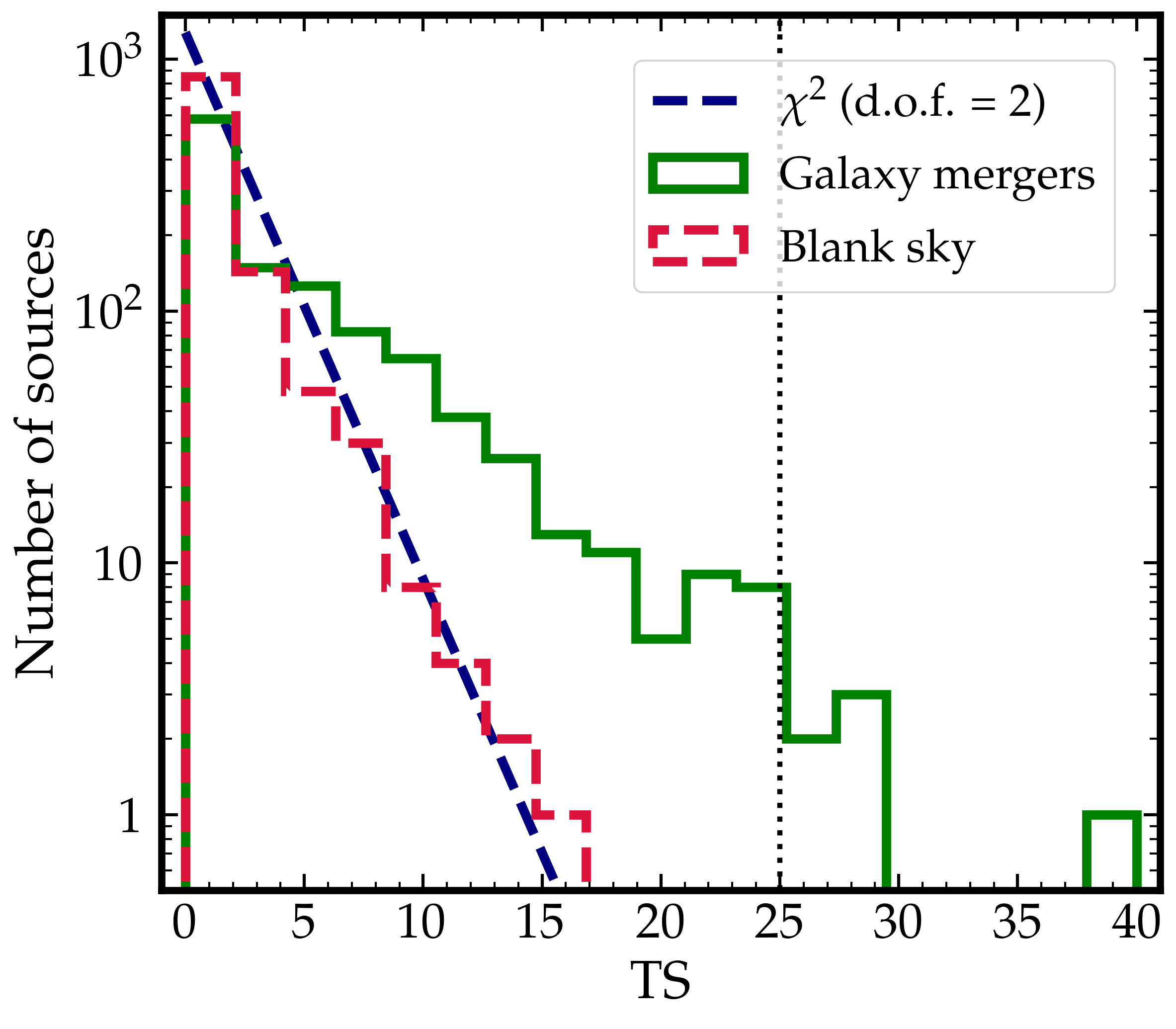}
    \caption{\justifying{TS distribution for 1,214 galaxy mergers (green solid), with the same number of randomly selected blank-sky regions (red dashed), and the chi-squared distribution with 2 degrees of freedom representing the null hypothesis (navy dashed). The vertical black dotted line marks $\rm TS=25$.}}
    \label{Fig:Histogram}
\end{figure}

\begin{table*}[ht]
\centering
\renewcommand{\arraystretch}{1.1}
\begin{threeparttable}
\vspace{0.3em}
{\footnotesize
\setlength{\tabcolsep}{5pt}
\renewcommand{\arraystretch}{1.5}  
\sisetup{separate-uncertainty=true, table-align-text-post=false}
\begin{tabular}{
    l                           
    c                           
    c                           
    S[table-format=1.2(2)]      
    S[table-format=1.2(3)e-2]   
    S[table-format=1.2(3)e-1]   
    S[table-format=2.2]         
}
\toprule
{Name} & {RA, DEC} & {RA, DEC} & {Spectral Index} & {Photon Flux} & {Energy Flux} & {TS} \\
& {[deg]} & {(loc) [deg]} & {($\Gamma$)} & {[ph cm$^{-2}$ s$^{-1}$]} & {[MeV cm$^{-2}$ s$^{-1}$]} & \\
\midrule
RCS2338 14991$^{a}$ & (358.21, $-11.58$) & (358.24, $-11.60$) & \num{1.62(17)} & \num{5.78(182)e-11} & \num{9.38(409)e-7} & 39.77 \\
GT1237664877803077714$^{b}$ & (144.21, 29.12) & (144.73, 29.23) & \num{2.78(35)} & \num{1.23(029)e-10} & \num{2.78(81)e-7} & 28.68 \\
GZ587742060519489665$^{c}$ & (144.97, 16.78) & (144.89, 16.83) & \num{2.58(31)} & \num{1.29(032)e-10} & \num{3.42(106)e-7} & 28.20 \\
RCS1645 10412$^{a}$ & (254.01, 38.85) & (253.77, 39.01) & \num{2.76(29)} & \num{1.63(037)e-10} & \num{3.74(99)e-7} & 27.52 \\
RCS1303 07302$^{a}$ & (193.33, $-5.06$) & (193.07, $-5.47$) & \num{2.77(16)} & \num{2.52(059)e-10} & \num{5.75(139)e-7} & 26.86 \\
RCS2329 14411$^{a}$ & (354.41, $-0.68$) & (354.33, $-0.56$) & \num{2.24(26)} & \num{1.04(034)e-10} & \num{4.17(148)e-7} & 26.50 \\
RCS2143 12345$^{a}$ & (327.88, 2.30) & (327.84, 2.38) & \num{1.70(20)} & \num{5.88(214)e-11} & \num{7.66(363)e-7} & 25.19 \\
GZ588017729228177602$^{c}$ & (191.78, 7.27) & (191.98, 7.19) & \num{1.90(21)} & \num{7.89(245)e-11} & \num{6.19(287)e-7} & 25.03 \\
\bottomrule
\end{tabular}}
\begin{tablenotes}
\footnotesize
\item $^{a}$~\cite{2009yCat..21810233H}; $^{b}$~\cite{2023MNRAS.524.4482B}; $^{c}$~\cite{2018MNRAS.479..415A}
\end{tablenotes}
\caption{Galaxy mergers detected in the 1–500 GeV range with significances $\gtrsim 5\sigma$ are shown with their catalog coordinates, \textit{Fermi}-LAT localised (loc) coordinates, spectral indices, photon fluxes, energy fluxes, and TS values. Superscripts refer to the sources' catalog references.}
\label{tab:SignificantGM}
\end{threeparttable}
\end{table*}

\begin{figure*}[!htbp]  
    \centering
    \includegraphics[width=0.48\textwidth]{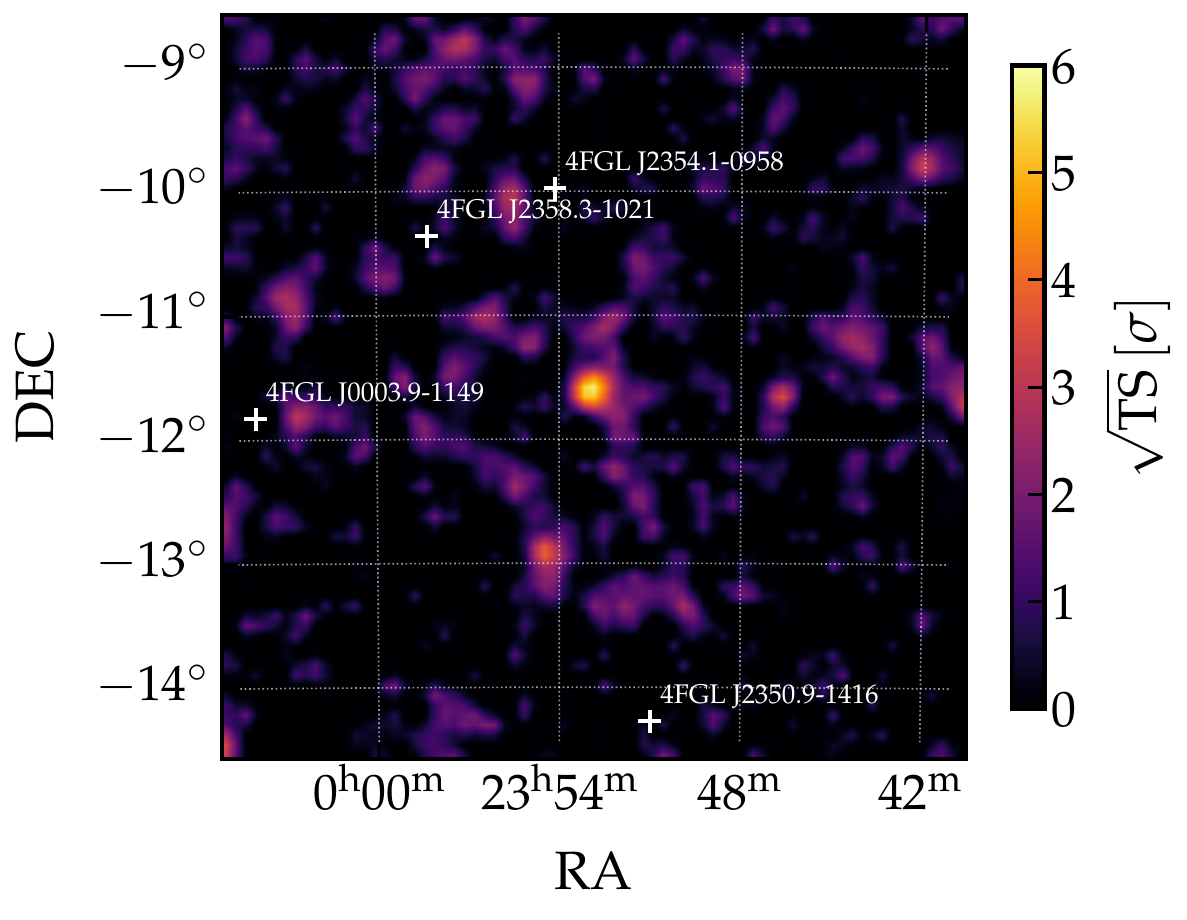}
    \hfill
    \includegraphics[width=0.48\textwidth]{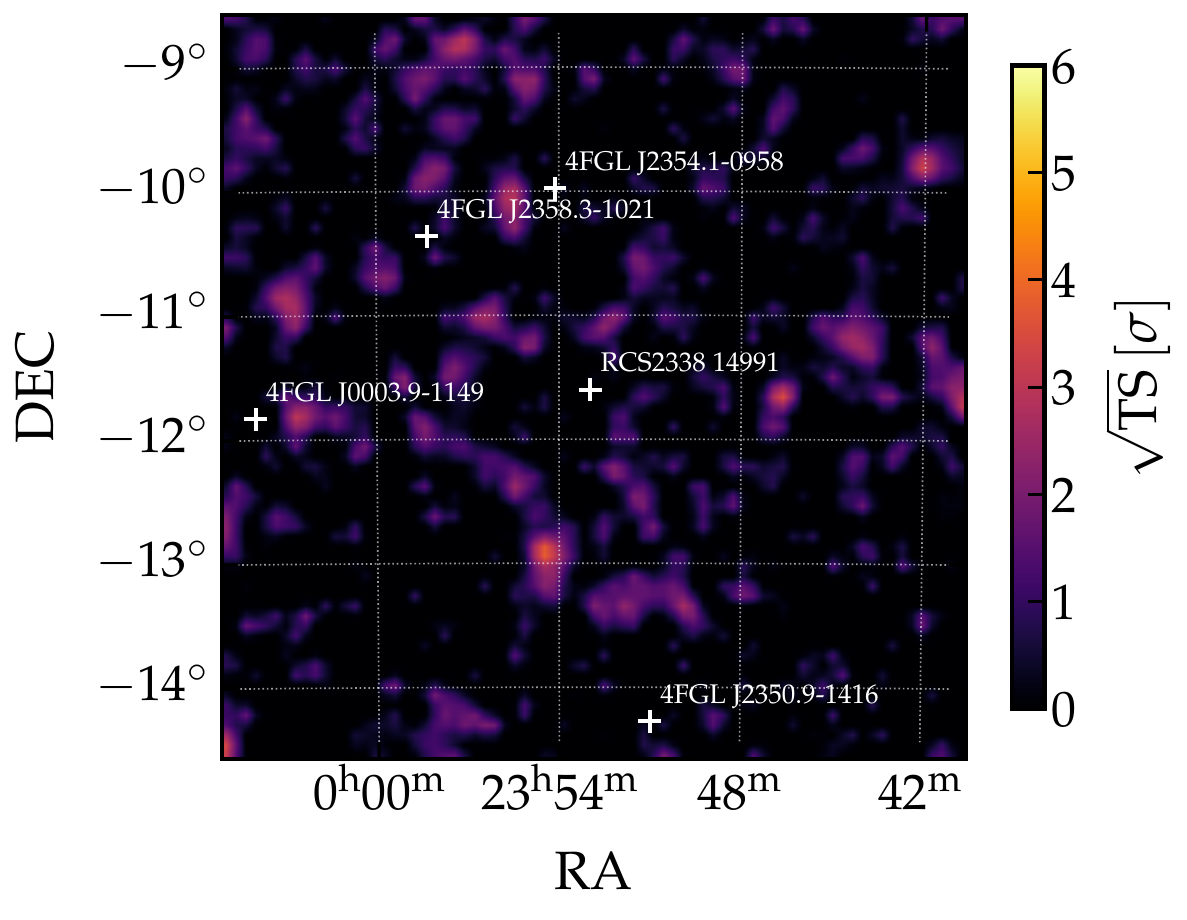}
    \caption{TS map centred on RCS2338 14991 for $\gamma$ rays in the energy range 1–500 GeV. The left panel shows the TS map without including the test source at the centre with a single \texttt{PowerLaw} spectrum. The right panel shows the TS map after inserting a test source at the centre. Both maps include existing 4FGL-DR4 catalog sources. }
    \label{fig:ts_map}
\end{figure*}

\begin{figure}[!htbp]
    \centering
    \includegraphics[width=1\linewidth]{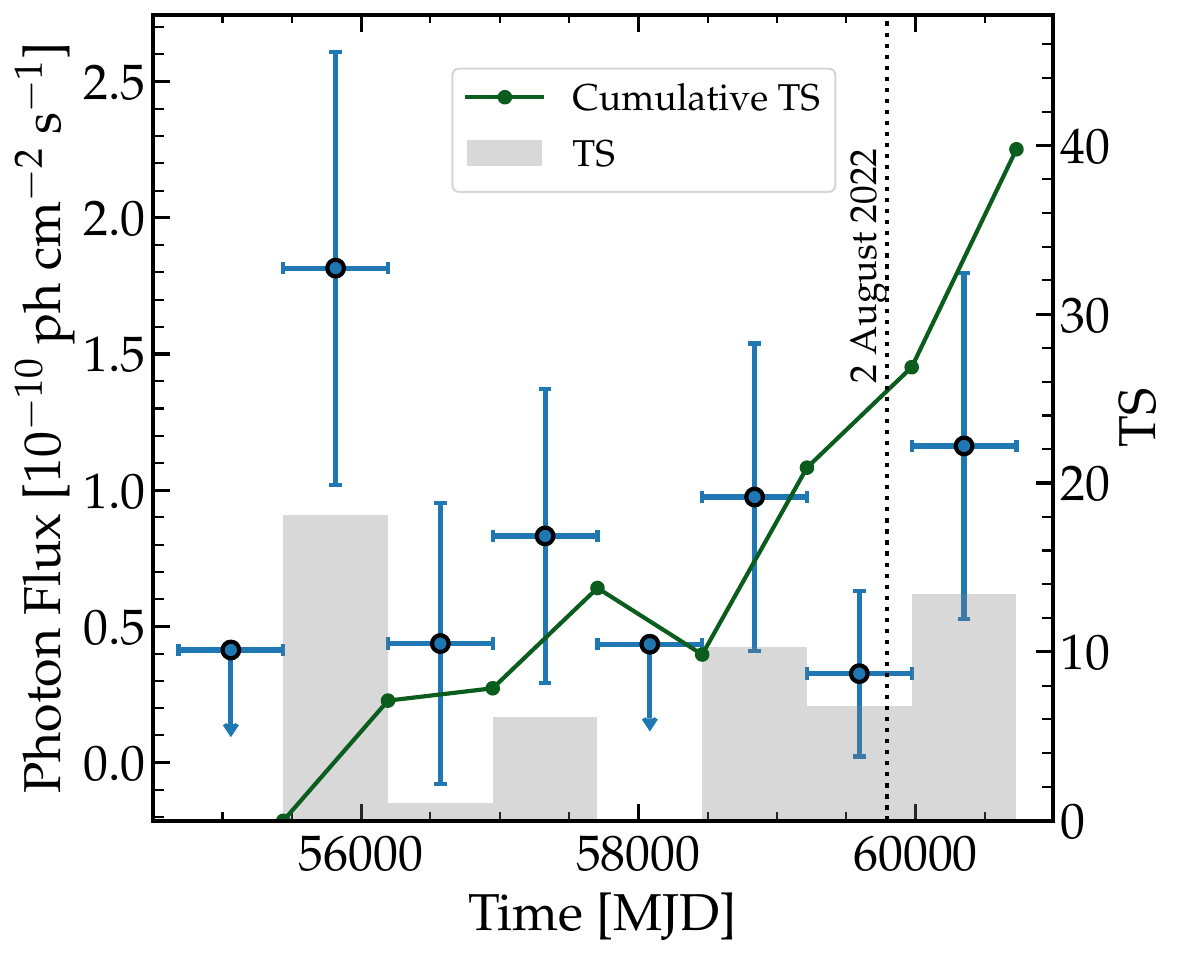}
    \caption{\justifying{Light curve of RCS2338 14991 spanning MJD 52697 to 60727. The blue points show the photon flux (1–500 GeV) in eight time bins, with error bars. Time bins with TS values very close to zero are shown with upper limits (95\% confidence level). The grey histogram displays the TS values for each time bin when TS is positive. The green curve represents the cumulative TS. The black dotted vertical line indicates the release date of 4FGL-DR4.}}
    \label{fig:lightcurve1}
\end{figure}

Here we present the results of individual analysis for the galaxy mergers considered in this study. We perform single-source analysis on the sample of 1,875 mergers, as described in Section~\ref{sec:Single-source-analysis}. We evaluate the quality of background modelling by retaining only those sources where both the galactic and extragalactic normalization values lie within 0.8–1.2, corresponding to a $20\%$ uncertainty \cite{Fermi-LAT:2010cni, Fermi-LAT:2017sxy}. Sources outside this range are excluded due to poor background modelling, which can arise from various factors such as contamination, incomplete background estimation, low flux from the target or complex source environments. This criterion yields 1,214 well-modelled galaxy mergers. 
In Fig.\,\ref{Fig:Histogram}, we show the TS distribution of these 1,214 well-modelled galaxy mergers and compare them with the same number of randomly selected blank-sky regions from a well-modelled pool of 1,700 blank-sky regions. The red dashed histogram represents the TS distribution of the blank-sky regions, which follows the $\chi^2$ distribution with 2 degrees of freedom (navy dashed line). As expected, the blank-sky regions are consistent with the statistical background fluctuations. In contrast, the green solid histogram shows the TS distribution of the considered galaxy mergers, which exhibit a number of sources with high TS values. Eight galaxy mergers show significant $\gamma$-ray emission with $\rm TS>25$. Table~\ref{tab:SignificantGM} lists these sources along with their analysis summary. RCS2338 14991 has the highest TS value of 39.77, corresponding to a detection significance of approximately $6 \sigma$. We also verified this result independently using the \texttt{Fermi ScienceTools} package and obtained consistent results. The remaining 1,206 galaxy mergers with $\rm TS<25$ are classified as faint sources. 

In Fig.\,\ref{fig:ts_map}, we show the $6\degree \times 6\degree$ TS map for the  ROI centered at RCS2338 14991. The left panel TS map is generated without including a test source model at the center, revealing excess emission at that location. The right panel shows the TS map after inserting a test source model at the center with a simple \texttt{PowerLaw} spectrum. We note that the new source model can fit the data well, leaving no excess emission. TS maps for the other seven sources are provided in Fig.\,\ref{fig:tsmaps_part2}.

\begin{table*}[!htbp]
\centering
\begin{threeparttable}
\sisetup{
    separate-uncertainty, 
    table-align-text-post=false 
}
\setlength{\tabcolsep}{7pt} 
\renewcommand{\arraystretch}{1.5}  
\begin{tabular}{
    l                               
    c                               
    S[table-format=1.2(3)e-2]       
    S[table-format=1.2(3)e-1]       
    S[table-format=1.2(2)]          
    S[table-format=2.2]             
}
\toprule
{Source Name} & {RA, Dec} & {Photon Flux} & {Energy Flux} & {Spectral Index} & {TS} \\
& {[deg]} & {[ph cm\(^{-2}\) s\(^{-1}\)]} & {[erg cm\(^{-2}\) s\(^{-1}\)]} & {(\(\Gamma\))} & \\
\midrule
NGC 3256        & {(156.96, \(-43.90\))} & \num{6.80(3.22)e-11} & \num{5.33(2.55)e-13} & \num{2.12(31)} & 12.00 \\
UGC 813/6         & {(19.07, 46.74)} & \num{5.12(2.81)e-11} & \num{4.31(2.27)e-13} & \num{2.08(33)} & 10.62 \\
NGC 660         & {(25.76, 13.65)} & \num{7.91(3.21)e-11} & \num{3.16(1.32)e-13} & \num{2.65(43)} &  9.97 \\ 
UGC 12914/5       & {(0.41, 23.48)} & \num{6.59(3.12)e-11} & \num{3.16(1.53)e-13} & \num{2.46(41)} &  7.96 \\ 
SDSS J0849+1114 & {(132.27, 11.25)} & \num{<7.48e-12} & \num{<1.75e-12} & \num{0.65(212)} &  6.36 \\
VV 114          & {(16.95, \(-17.51\))} & \num{<3.35e-11} & \num{<1.75e-13} & \num{2.39(922)} &  0.09 \\ 
ESO 303-11       & {(64.91, \(-39.17\))} & \num{<3.25e-11} & \num{<9.07e-14} & \num{3.35(1800)} &  0.00 \\
\bottomrule
\end{tabular}
\end{threeparttable}
\caption{Analysis summary of the theoretically motivated galaxy mergers \cite{Kashiyama:2014rza,Yuan:2018erh,Yuan:2017dle}. Sources whose flux is consistent with zero are listed with the corresponding 95\% confidence level upper limits. }
\label{tab:th_galaxies}
\end{table*}

To assess the robustness of the significant signal from RCS2338 14991, we perform an analysis over the considered period divided into two intervals: August 4, 2008, to August 2, 2022, corresponding to the observational time of the 4FGL-DR4 catalog \cite{Ballet:2023qzs}, and August 2, 2022, to February 21, 2025, covering the period after 4FGL-DR4. We find TS values of 18.30 and 21.80 for the first and second intervals, respectively. The TS value below the catalog threshold during the 4FGL-DR4 period explains why the source is not included in the \textit{Fermi}-LAT catalog. Furthermore, we show the photon flux light curve for this highly significant source in Fig.\,\ref{fig:lightcurve1}, where the blue data points show the photon flux in each time bin with uncertainties in the flux, the grey bars represent the associated TS value, and green points show the cumulative TS. 

\begin{table}[htbp]
\centering
\normalsize
\setlength{\tabcolsep}{12pt} 
\begin{tabular}{
    c                               
    c                               
}
\toprule
{Photon Energy} & {Event Time}\\
{[GeV]} & {[yyyy-mm-dd]}\\
\midrule
112.751 & 2023-12-23 \\ \addlinespace
83.217 & 2022-11-30 \\ \addlinespace
23.940 & 2012-03-07 \\ \addlinespace
13.101 & 2020-11-11 \\ \addlinespace
\bottomrule
\end{tabular}
\caption{\justifying{Photon Events from the direction of RCS2338 14991 with probability $> 0.95$ in the energy range 1--500 GeV.}}
\label{tab:ph_events}
\end{table}

Additionally, we utilize \texttt{gta.compute\_srcprob} to compute the probability of the events detected in the direction of RCS2338 14991 ROI. For this, we use the model parameters obtained from the likelihood fitting. Table~\ref{tab:ph_events} lists the photon events whose probability is $>0.95$, listing their corresponding energies and detection times. Notably, two HE photons (112.751 GeV and 83.217 GeV) were detected after the 4FGL-DR4 catalog release, contributing significantly to the increased TS value observed for this source, as discussed above.

Since RCS2338 14991 is situated within a galaxy cluster~\cite{2009yCat..21810233H}, there may be additional galaxy mergers in its vicinity that \textit{Fermi}-LAT cannot resolve. According to  Table~\ref{tab:ph_events}, there are two photon events of energy near 100 GeV with probabilities $>$ 0.95, which may significantly boost the TS. Given that the \textit{Fermi}-LAT PSF at 100 GeV is $\lesssim$ 0.1$\degree$, we identified two/\,four galaxy mergers within 0.05$\degree$/\,0.1$\degree$ radii. These sources were excluded from our final sample list based on selection criteria as described in Section~\ref{sec:sample cut}. To clarify the origin of the emission, we perform the individual source analysis on these four neighboring mergers and find that two of them, RCS2338 14994 (at 0.02$\degree$) and RSC2338 15001 (at 0.04$\degree$), yield similar results as RCS2338 14991 mentioned in Table\,\ref{tab:SignificantGM}. Therefore, it is possible that the HE $\gamma$-ray events mentioned in Table~\ref{tab:ph_events} originate from either of these two neighbouring sources.

\begin{figure*}[!htbp]  
    \centering
    \includegraphics[width=0.48\textwidth]{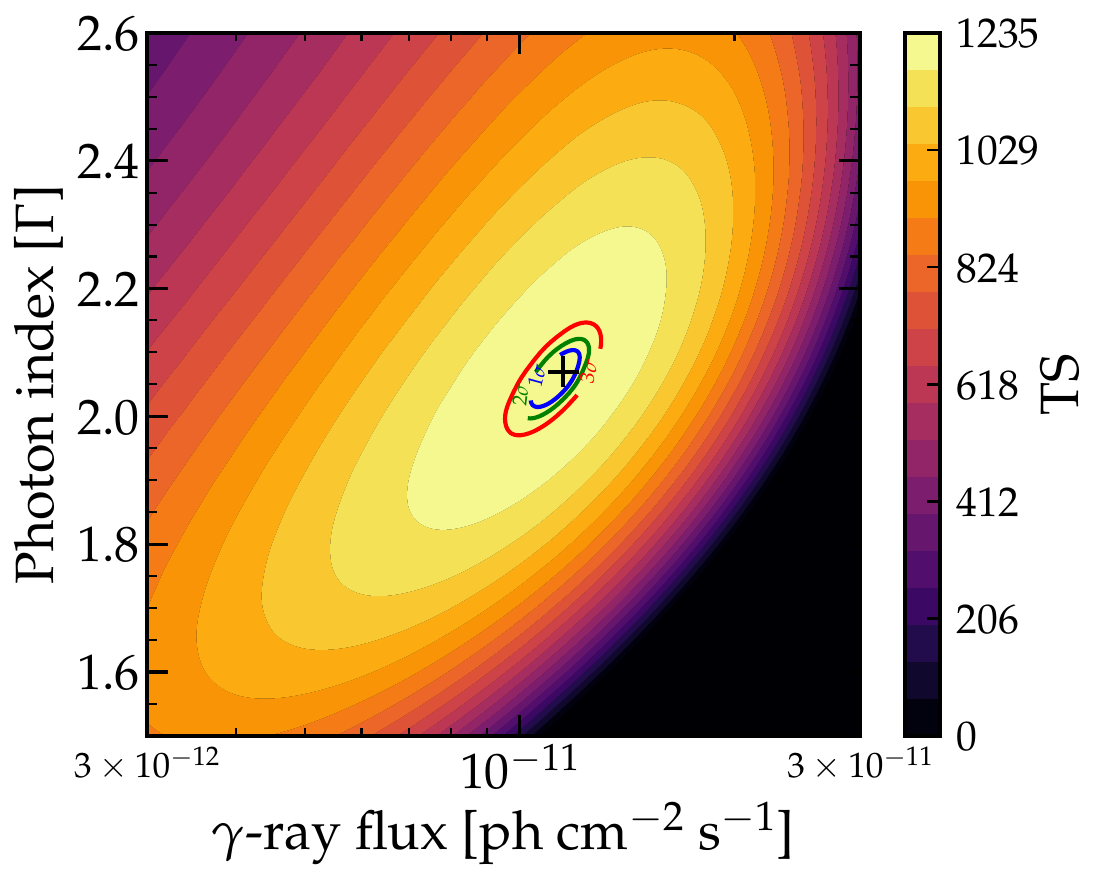}
    \hfill
    \includegraphics[width=0.48\textwidth]{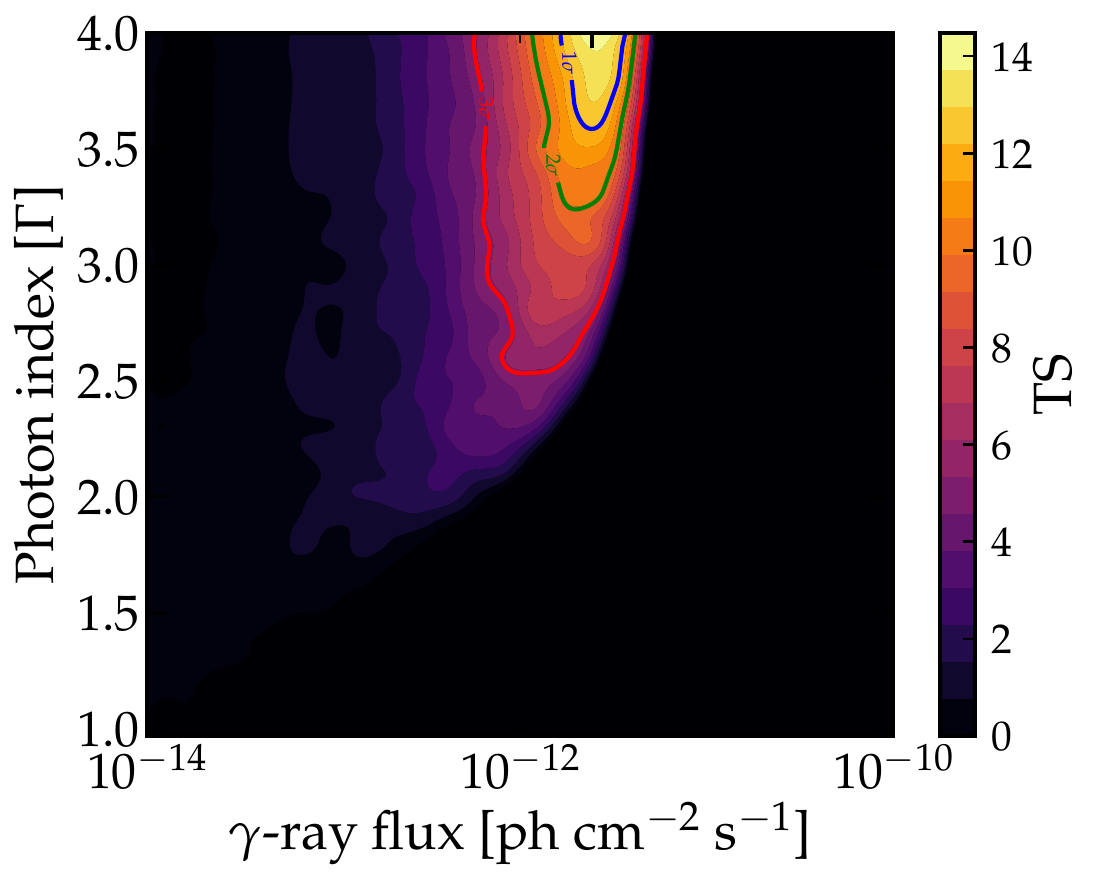}
    \caption{\justifying{\textit{Left}:\,\,Stacked TS profile of 1,206 faint galaxy mergers (TS $<$ 25) with best-fit photon index and flux represented by a black cross. Blue, green, and red contours represent $1\sigma$, $2\sigma$, and $3\sigma$ confidence levels, respectively. The color bar represents the stacked TS values. \textit{Right}:\,\,Stacked TS profile of 1,206 randomly selected blank-sky regions. }}
    \label{fig:ts_combined}
\end{figure*}

\subsection*{Theoretically motivated galaxy mergers}
\label{subsec: Theoretically_motivated_GM}
Beyond analyzing galaxy merger samples from survey catalogs, we also investigate galaxy mergers that have been theoretically motivated as potential sources of HE $\gamma$-ray emission in Refs.\,\cite{Kashiyama:2014rza, Yuan:2017dle, Yuan:2018erh}. None of these objects fall within the 95\% confidence uncertainty region of any 4FGL-DR4 source. We therefore carry out dedicated source-by-source analyses for these targets and summarize the results in Table \ref{tab:th_galaxies}. We observe a weak signal from NGC 3256 ($\sim 3.46\sigma$), UGC 813/6 ($\sim 3.26\sigma$), NGC 660 ($\sim 3.16\sigma$), and UGC 12914/5 ($\sim 2.82\sigma$). For sources with flux values consistent with zero, we report the corresponding upper limits in Table~\ref{tab:th_galaxies}. We note that flux from NGC 3256 and NGC 660 is consistent with the calculated $\gamma$-ray flux of $\mathcal{O}(10^{-13}\rm\ erg\,cm^{-2}\,s^{-1})$ reported in Ref.\,\cite{Yuan:2018erh}. A recent analysis by \cite{Manna:2025eob} considered a subset of these sources. Our results are consistent with theirs for NGC 3256, NGC 660, and VV 114, though they did not observe marginal significance from UGC 813/6 or UGC 12914/5.

\subsection{Stacked Galaxy Merger Analysis}
\label{sec:stacked-results}

Here we present the results of stacking the faint galaxy mergers. The stacking analysis is performed on 1,206 faint galaxy mergers, following the procedure described in Section~\ref{sec:stacking analysis}. To ensure that the analysis captures the collective behavior of faint sources, high--significance ($\rm TS>25$) sources are excluded. The left panel of Fig.\,\ref{fig:ts_combined} shows the 2D stacked TS profile for these faint galaxy mergers. We perform a 2D \texttt{RectBivariateSpline} fit on the discrete photon index-flux surface to get a smooth continuous TS surface~\cite{2020SciPy-NMeth}. It helps to extract the accurate best-fit parameters and robust confidence contours by suppressing numerical fluctuations between grid points. The best-fit parameters--spectral index $\Gamma = 2.07^{+0.27}_{-0.16}$ and photon flux of $\rm (1.15^{+0.42}_{-0.73} \times 10^{-11})~ph~cm^{-2}~s^{-1}$--indicated by a black cross, yield a peak stacked TS of 1235, corresponding to a detection significance of approximately $\sim35\sigma$ for 2 degrees of freedom. Confidence contours of $1\sigma,~2\sigma$ and $3\sigma$ are shown by blue, green and red contours, respectively. We note that according to Ref.\,\cite{Kashiyama:2014rza}, galaxy mergers can accelerate CRs up to ultra-high energies (UHEs) with a hard spectral index of $\Gamma \approx 2.2$ and Ref.\,\cite{Yuan:2017dle} predicts that the CR spectrum produced by galaxy mergers would have spectral indices in the range $1.5 \lesssim \Gamma \lesssim 2.0$. Our findings are in close agreement with these theoretical expectations.

To validate the background modelling, we stack an equal number of randomly selected blank-sky regions. The right panel of Fig.\,\ref{fig:ts_combined} presents these results, showing a maximum collective $\rm TS \sim 15$ peaks at a spectral index of 4.0 and a photon flux of $\rm \sim 2\times10^{-12}~ph~cm^{-2}~s^{-1}$. The peak of low TS corresponds to a much softer spectral index than that of the galaxy mergers, indicates a negligible contribution of random background fluctuations to the stacked $\gamma$-ray emission from galaxy mergers. Therefore, our findings confirm that this stacking method can effectively unveil the cumulative HE $\gamma$-ray emission from an entire population of faint sources.

\subsection*{Stacked SED}
\label{subsec:SED}
\begin{figure}
    \centering
    \includegraphics[width=1\linewidth]{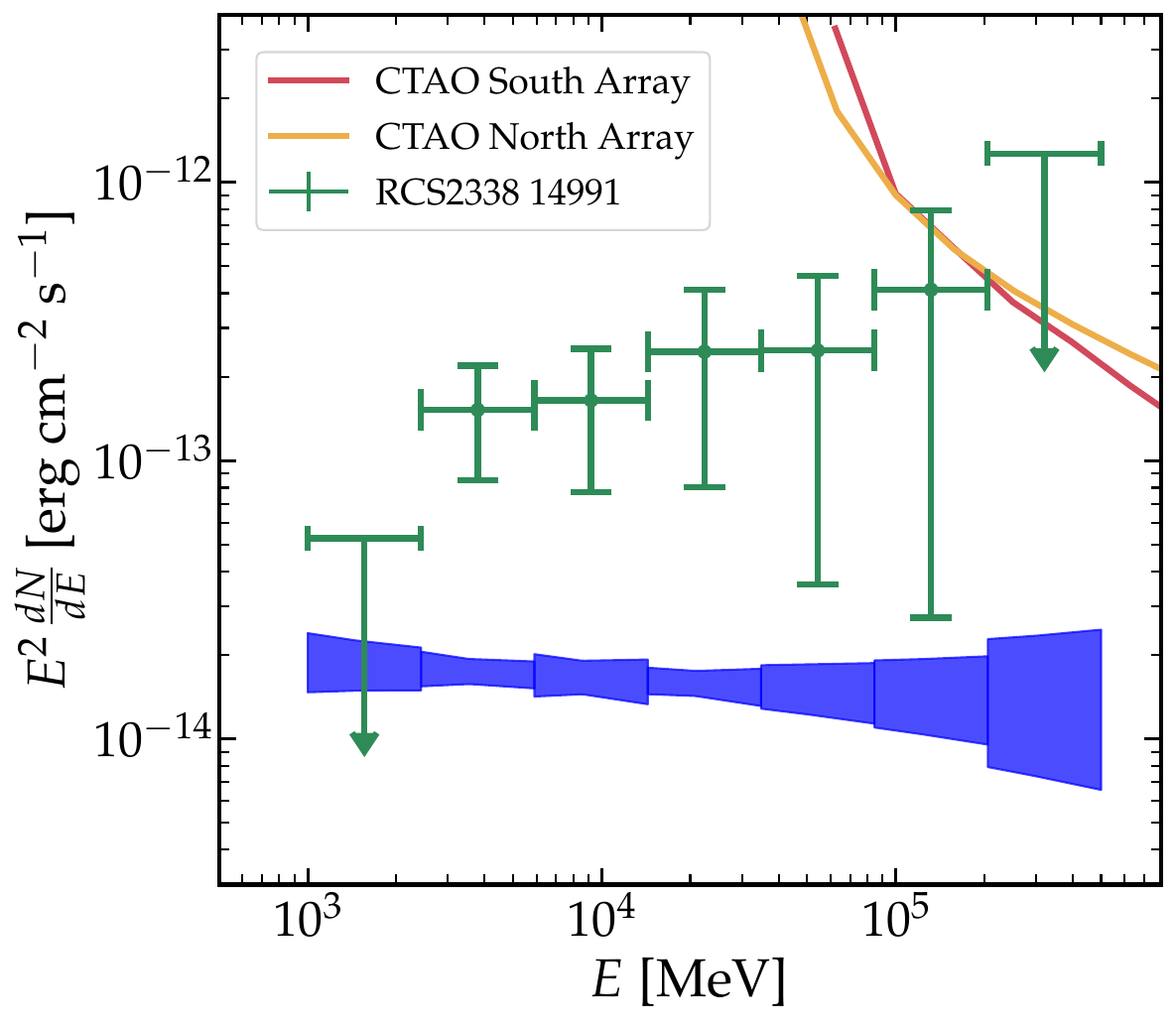}
    \caption{\justifying{ Gamma-ray stacked SED of 1206 galaxy mergers with $\rm TS < 25$. The spectrum is shown as a bow-tie plot with a blue shaded region in seven logarithmically spaced energy bins between 1 GeV and 500 GeV. For comparison, the SED of the most significant source, RCS2338 14991, is shown with green points and error bars. The differential sensitivity curves for 5$\sigma$ detection of point sources with 50 hr observation time are shown for CTAO Southern array (red solid curve) and CTAO Northern array (orange solid curve).}}
    \label{fig:placeholder}
\end{figure}
We present the stacked SED of the faint galaxy mergers analyzed in this study. The stacking of these sources in the full energy range of 1–500 GeV yields a highly significant $\gamma$-ray detection at the $\sim35\sigma$ level, enabling us to examine the spectral characteristics of the stacked signal in smaller energy intervals. We divide the 1–500 GeV range into seven logarithmically spaced bins and perform the same stacking analysis in each bin to determine the best-fit spectral index and photon flux. We found all seven bins exhibit high stacked TS values, allowing us to construct the corresponding butterfly plots, as shown in Fig.\,\ref{fig:placeholder}. Notably, the last energy bin (205--500 GeV) also exhibits a high stacked TS value, leading us to further investigation. We identified 28 photon events above 200 GeV from the sample of 1,206 faint galaxy mergers with event probability $\gtrsim 90\%$ listed in Table\,\ref{tab:200GeV}. The presence of these HE photon events explains the significant TS observed in the last energy bin of the SED. Moreover, four of these events have energies greater than 400 GeV with the most energetic event with 475\,GeV , making them compelling candidates for targeted TeV searches in the direction of the corresponding galaxy mergers. There are no TeVCat sources found within 1\degree\ of these galaxy mergers~\cite{2008ICRC....3.1341W}. The spectral index in all bins is found to be $\Gamma \sim 2.07$, consistent with theoretical expectations mentioned in Refs.\,\cite{Kashiyama:2014rza, Yuan:2017dle, Yuan:2018erh,2011arXiv1104.0956V}. The stacked spectral energy flux in each bin is approximately $\rm 2\times 10^{-14}~erg~cm^{-2}~s^{-1}$, which is close to theoretically calculated $\gamma$-ray flux of $\mathcal{O}(10^{-13}-10^{-14} \,\rm erg\,cm^{-2}\,s^{-1})$ in Refs.\,\cite{Kashiyama:2014rza,Yuan:2017dle,Yuan:2018erh} and few times higher compared to the flux of $\mathcal{O}\rm (10^{-15}~erg~cm^{-2}~s^{-1})$ predicted for integral hadronic and inverse Compton emission of HE $\gamma$-ray photons in Ref.\,\cite{2011arXiv1104.0956V}, respectively, from mergers located at $\mathcal{O}(50~\rm Mpc)$. We also present the SED of RCS2338 14991 in Fig.\,\ref{fig:placeholder}, shown as green data points with associated energy flux uncertainties. For comparison, we overlay the $5\sigma$ detection sensitivity limits of the upcoming Cherenkov Telescope Array Observatory (CTAO) Southern array (red solid line) and Northern array (orange solid line) for a 50-hour exposure.

\subsection{Galaxy Merger Association with 4FGL-DR4 sources}
\label{subsec:unassociated}
The most recent 4FGL–DR4 catalog reports 7,195 point sources detected over 14 years of \textit{Fermi}-LAT observations in the energy range 50 MeV to 1 TeV \cite{Ballet:2023qzs}, of which 2,339 remain unassociated\footnote{Sources without an identified counterpart, as listed in \texttt{gll\_psc\_v35}}. These unassociated sources lack identified counterparts at other wavelengths and are detected only in $\gamma$ rays. We cross-match the sky coordinates of these sources with those of 31,464 galaxy mergers in our full sample and find that 18 unassociated sources have at least one galaxy merger within their 95\% uncertainty region. Table\,\ref{tab:unassocia} contains the list of these unassociated 4FGL-DR4 sources with their corresponding galaxy mergers and the separation between them. A recent study by \cite{Manna:2025bqo} investigated spatial coincidences between all \textit{Fermi}-LAT sources and galaxy mergers drawn from \href{https://cdsarc.cds.unistra.fr/viz-bin/cat/J/MNRAS/479/415}{J/MNRAS/479/415} and \href{https://cdsarc.cds.unistra.fr/viz-bin/cat/J/AJ/130/2043}{J/AJ/130/2043} catalogs. They reported 21 statistically significant associations within the $4\sigma$ positional uncertainty threshold ($p < 0.05$). In contrast, we focus specifically on associations with \textit{Fermi}-unassociated sources.

\section{Discussion and Conclusion}
\label{sec:discussion}

For the first time, this study systematically searches for the possibility that the dynamical system of galaxy mergers is HE $\gamma$-ray emitters utilizing 16.7 years of \textit{Fermi}-LAT data. Starting with an astronomical 31,464 galaxy merger sources from eight different catalogs and carefully reducing the number to minimize the $\gamma$-ray contamination from the bright $\gamma$-ray sources in the sky, we provide compelling evidence that galaxy mergers are HE $\gamma$-ray emitters. We show that galaxy mergers contribute to the extragalactic diffuse $\gamma$-ray sky. 

We follow two lines of evidence. Our individual source analysis shows that 8 galaxy mergers exhibit $\gtrsim5\sigma$ detection of HE $\gamma$-ray emission. The most prominent one, RCS2338 14991, is detected with a near $6\sigma$ significance with a photon flux of $\rm 5.78 \times 10^{-11}~ph~cm^{-2}~s^{-1}$ in the 1--500 GeV range. We have found a ~112 GeV photon associated with this source with more than 95\% CL. We also discuss why this source was not included in the 4FGL-DR4 catalog, as it was less significant at the time of its publication. This shows that at least a small subset of these mergers is detectable as point sources in the $\gamma$-ray sky. With the future CTAO telescopes, we hope to see more galaxy mergers to be detected as HE $\gamma$-ray sources.

Secondly, realizing that most galaxy mergers are too faint to be detected as point sources, we conduct a stacking analysis with these dim sources. We find a highly significant collective signal with TS of 1235 corresponding to a detection of $\sim35 \sigma$ significance. We assume the spectrum of the stacked population is a \texttt{PowerLaw2}. We find the best fit spectral index is $\sim$ 2.07 and an average energy flux of $\rm \sim 2 \times 10^{-14}~erg~cm^{-2}~s^{-1}$. The best fit spectral index $\Gamma$ is in good agreement with theoretical studies. Moreover, the energy flux is consistent with the theoretical inverse Compton and hadronic emission models. The observed $\gamma$ rays possibly originate through the inverse Compton scattering of the ambient photons by the electron-positron pairs, which are produced when VHE $\gamma$ rays annihilate with low-energy photons. Furthermore, our null results from the blank sky stacked analysis suggest that our findings are robust and are not statistical fluctuations. 

We identify 28 HE photon events with energies exceeding 200\,GeV from the sample of faint galaxy mergers. Among them, four events have energies greater than 400\,GeV, with the most energetic photon of 475.39\,GeV from RCS2143 12477. Since none of these galaxy mergers coincide with any known Fermi 4FGL-DR4 or TeVCat sources, these galaxy mergers emerge as candidates for targeted TeV searches.

Additionally, from our cross-check study, we show that galaxy mergers may be the counterparts of at least 18 unassociated $\gamma$-ray sources listed in the 4FGL-DR4 catalog. 

The implications of our findings are diverse. Our results establish that galaxy mergers are HE $\gamma$-ray emitters and a part of the extragalactic diffuse $\gamma$-ray sky. It paves a path to multi-messenger astronomy to understand the VHE astrophysical sites of galaxy mergers and their connection to CR physics and neutrino astronomy. 

\vspace{1.5em}
\textbf{\textit{Note added.}} While this work was in progress, two related studies from the same group appeared on arXiv~\citep{Manna:2025bqo, Manna:2025eob}. These works overlap with some aspects of our analysis with a limited set of galaxy mergers. 
\vspace{1.5em}
\section*{Acknowledgements} 
We thank Pugazhendi A D, Sushmita Agarwal, Debajit Bose, Anirban Das, Christopher Karwin, Kazumi Kashiyama, Vaidehi Paliya, Akash Kumar Saha, and Tracy Slatyer for various discussions and helpful comments. This research work has utilized data from NASA's Fermi Gamma-ray Telescope Support Center. We acknowledge the Fermi Collaboration for providing open access to the \faGithub\ \href{https://github.com/ckarwin/Fermi\_Stacking\_Analysis}{stacking analysis code}\,\cite{karwinFermistacking}, which facilitated our stacking analysis. We acknowledge the use
of high-performance computational facilities at the PTG Cluster at the Department of Physics, Indian Institute of Science, Bengaluru, India. S.B. acknowledges the Council of Scientific and Industrial Research (CSIR), Government of India, for supporting his research under the CSIR Junior/Senior Research Fellowship program through grant no. 09/0079(15488)/2022-EMR-I. P.B. acknowledges support from the COFUND action of Horizon Europe’s Marie Sklodowska-Curie Actions research programme, Grant Agreement 101081355 (SMASH). M.D. acknowledges the support of the Department of Science and Technology (DST) grant
DST/WIDUSHIA/PM/2023/25(G) for this research. R.L.\,\,acknowledges financial support from the institute start-up funds and ISRO-IISc STC for the grant no. ISTC/PHY/RL/499.

\bibliographystyle{JHEP}
\bibliography{ref.bib}

\clearpage
\appendix
\onecolumngrid
\setcounter{figure}{0}
\setcounter{table}{0}
\setcounter{equation}{0}

\renewcommand{\thefigure}{A\arabic{figure}}
\renewcommand{\thetable}{A\arabic{table}}
\renewcommand{\theequation}{A\arabic{equation}}

\section{TS maps of significant Galaxy Mergers}
\label{sec:Appendix A}

Here, we present $6\degree \times 6\degree$ TS maps for all the significant galaxy mergers with $\mathrm {TS>25}$ mentioned in Table\, \ref{tab:SignificantGM}. See Fig.\,\ref{fig:ts_map} for the TS maps of RCS2338 14991. In Fig.\,\ref{fig:tsmaps_part2}, the left panel for each row represents the case where the test source is not included in the model fitting. We observe a residual hot-spot at the center of these ROIs. The right panel describes the case where we include the source with a \texttt{PowerLaw} spectrum in the model fitting. We note that inclusion of the test source explain the excess of the $\gamma$-ray in the TS maps.  

\begin{figure*}[!htbp]
    \centering
    \begin{subfigure}{0.48\textwidth}
        \includegraphics[width=\linewidth]{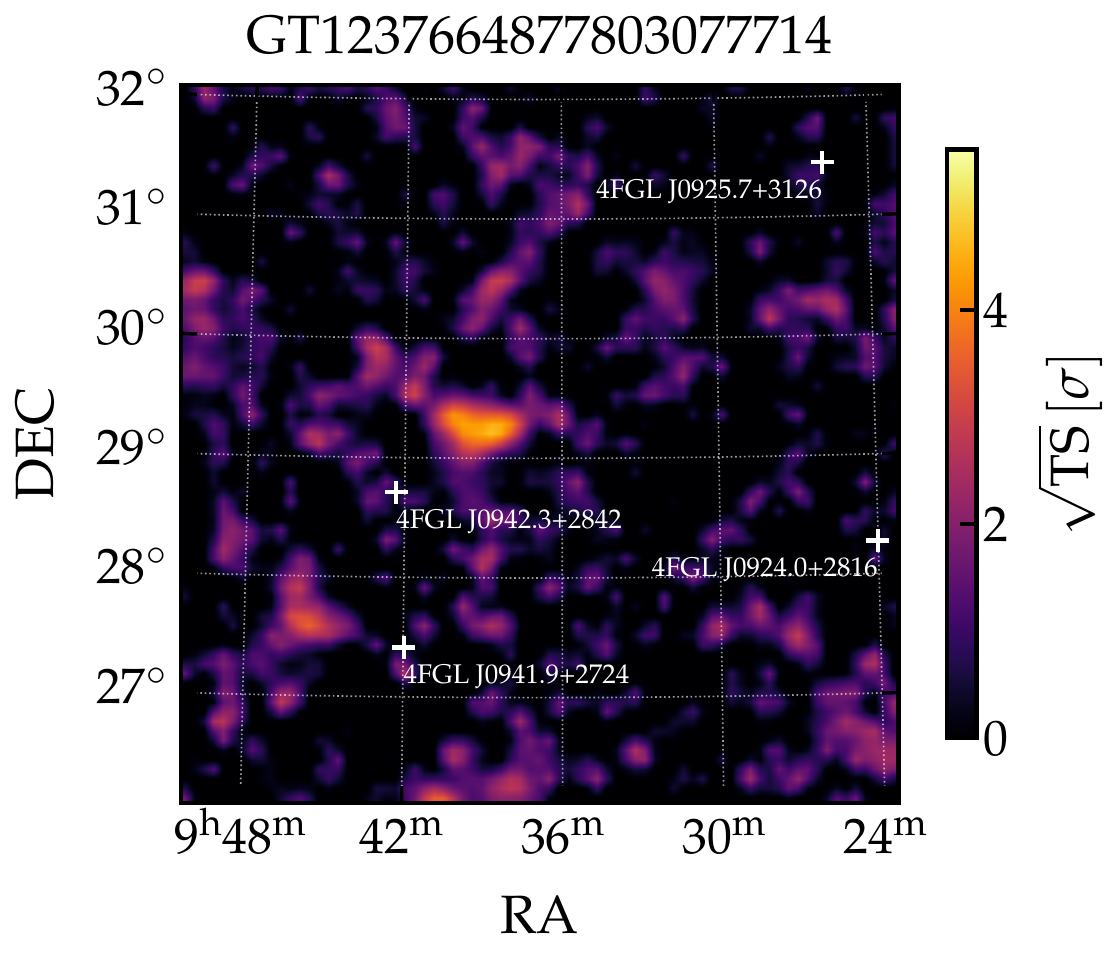}
    \end{subfigure}
    \hfill
    \begin{subfigure}{0.48\textwidth}
        \includegraphics[width=\linewidth]{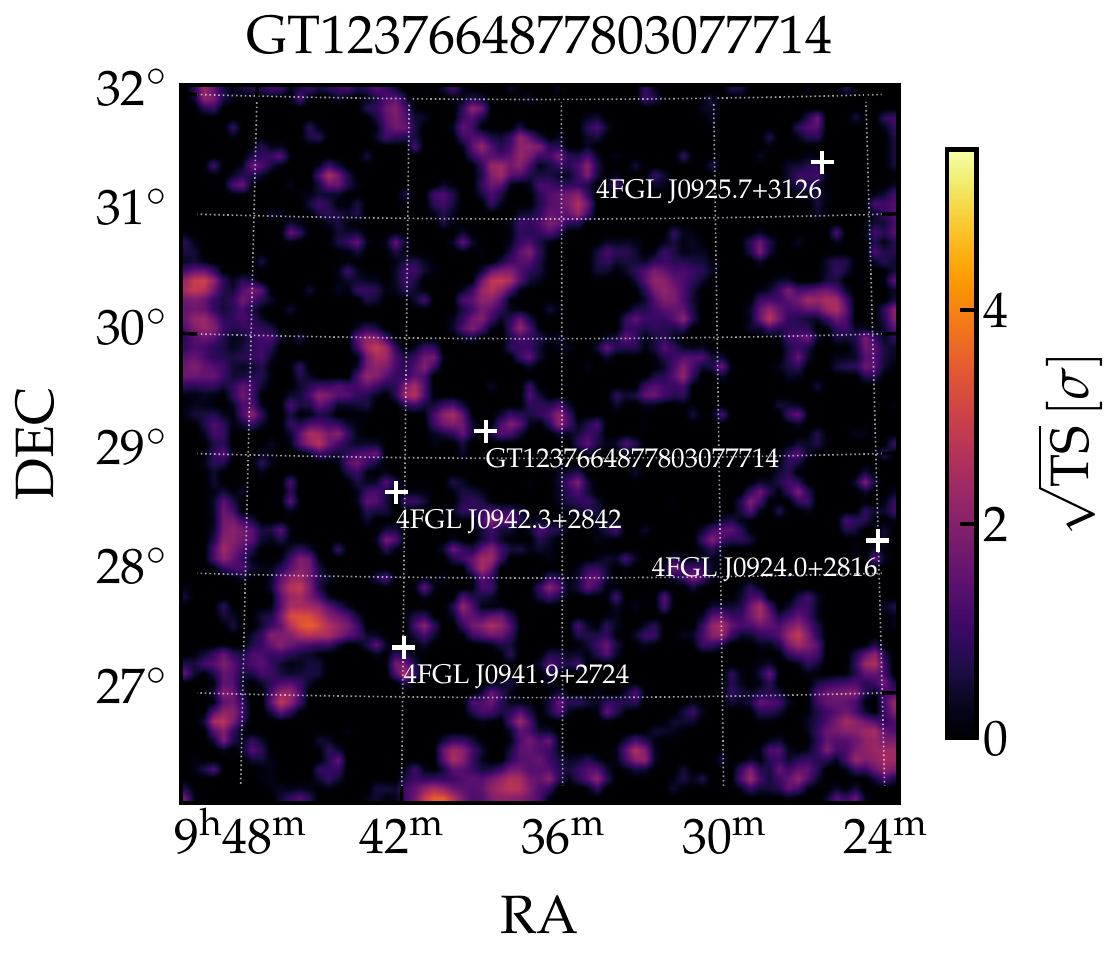}
    \end{subfigure}
\end{figure*}

\begin{figure*}[!htbp]
    \ContinuedFloat
    \centering
    \begin{subfigure}{0.48\textwidth}
        \includegraphics[width=\linewidth]{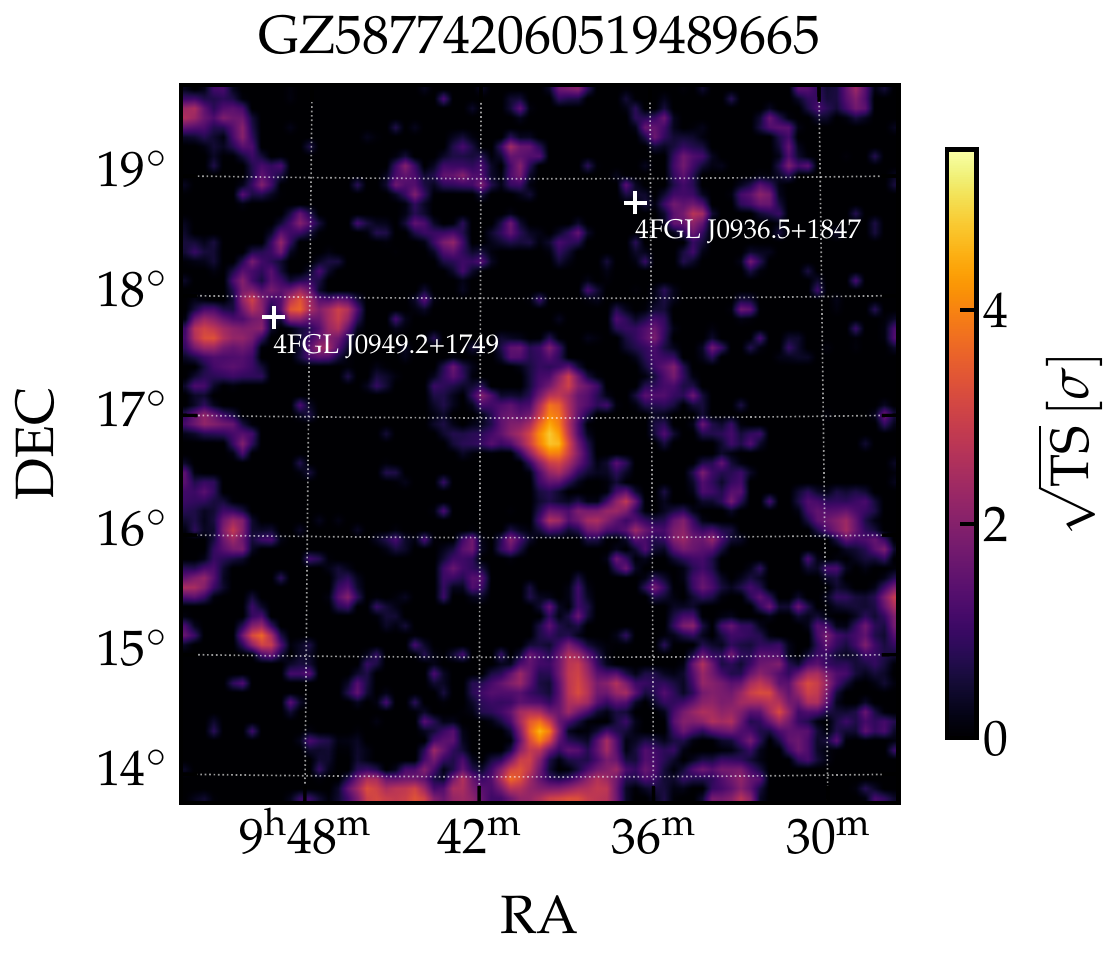}
     
    \end{subfigure}
    \hfill
    \begin{subfigure}{0.48\textwidth}
        \includegraphics[width=\linewidth]{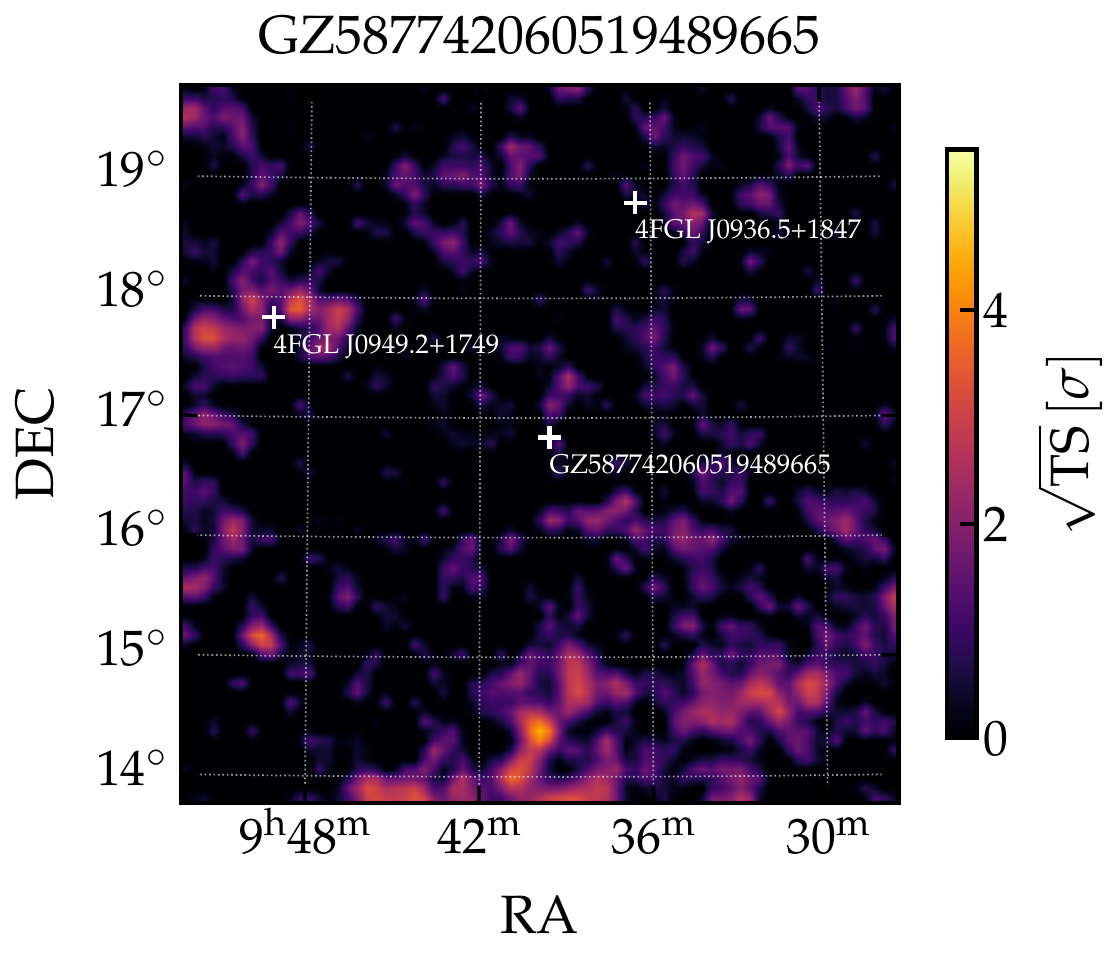}
        
    \end{subfigure}
\end{figure*}

\begin{figure*}[!htbp]
    \ContinuedFloat
    \centering    
    \begin{subfigure}{0.48\textwidth}
        \includegraphics[width=\linewidth]{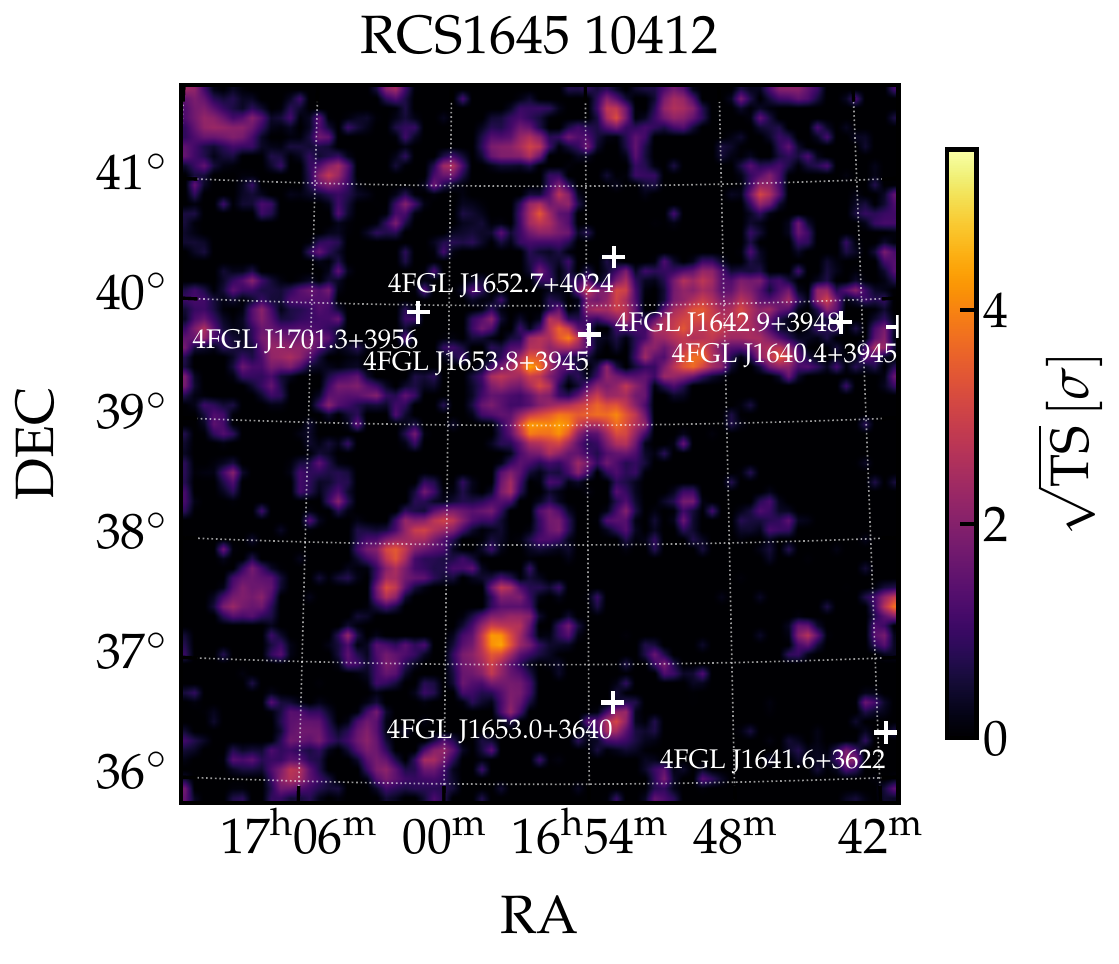}
       
    \end{subfigure}
    \hfill
    \begin{subfigure}{0.48\textwidth}
        \includegraphics[width=\linewidth]{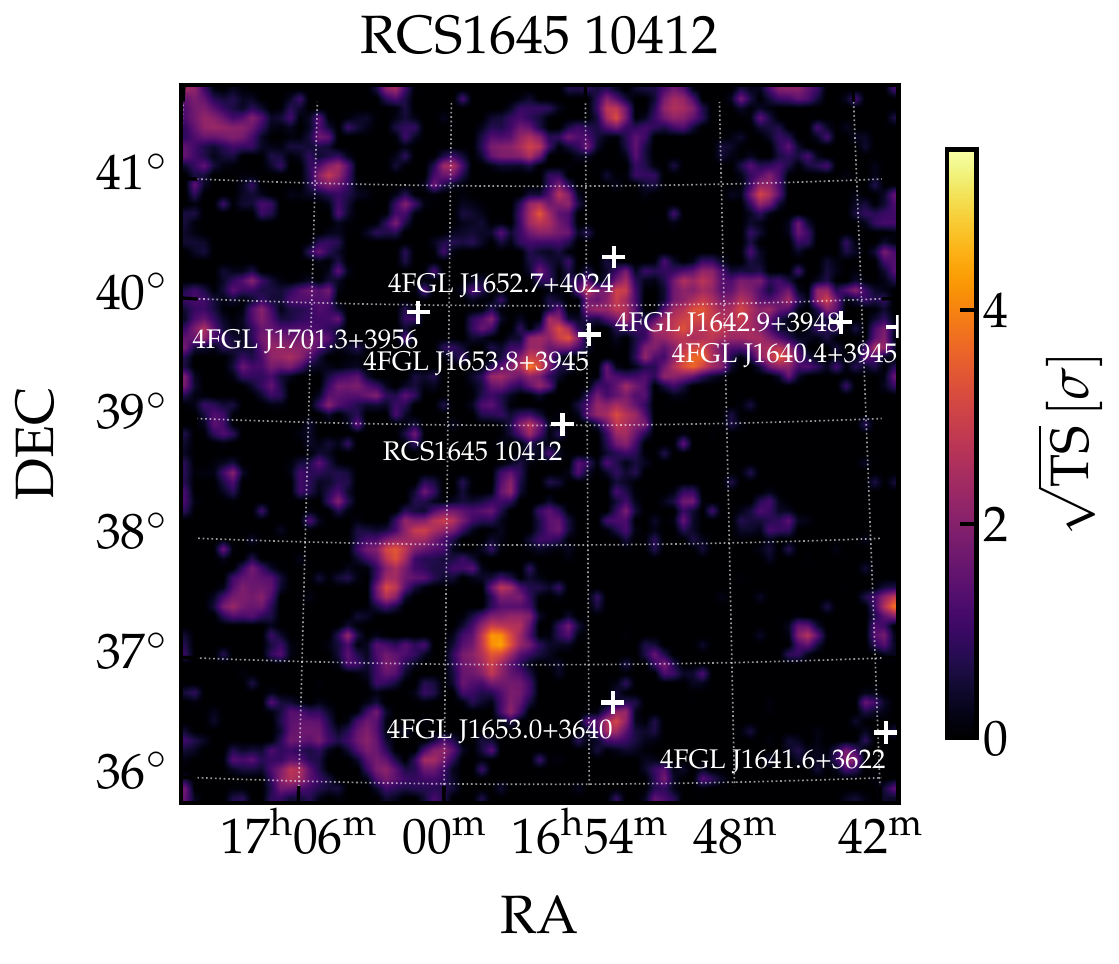}
       
    \end{subfigure}
\end{figure*}

\begin{figure*}[!htbp]
    \ContinuedFloat
    \centering
    \begin{subfigure}{0.48\textwidth}
        \includegraphics[width=\linewidth]{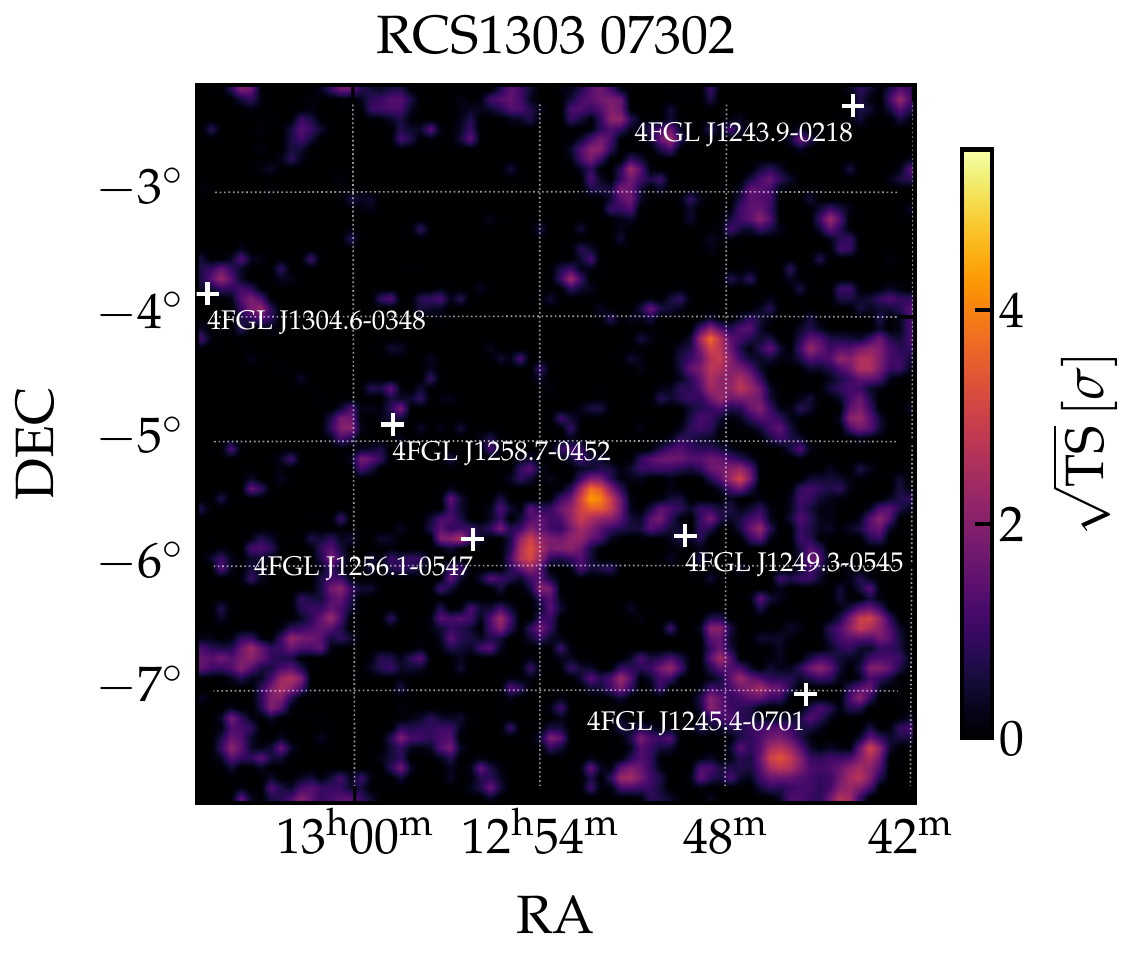}
    \end{subfigure}
    \hfill
    \begin{subfigure}{0.48\textwidth}
        \includegraphics[width=\linewidth]{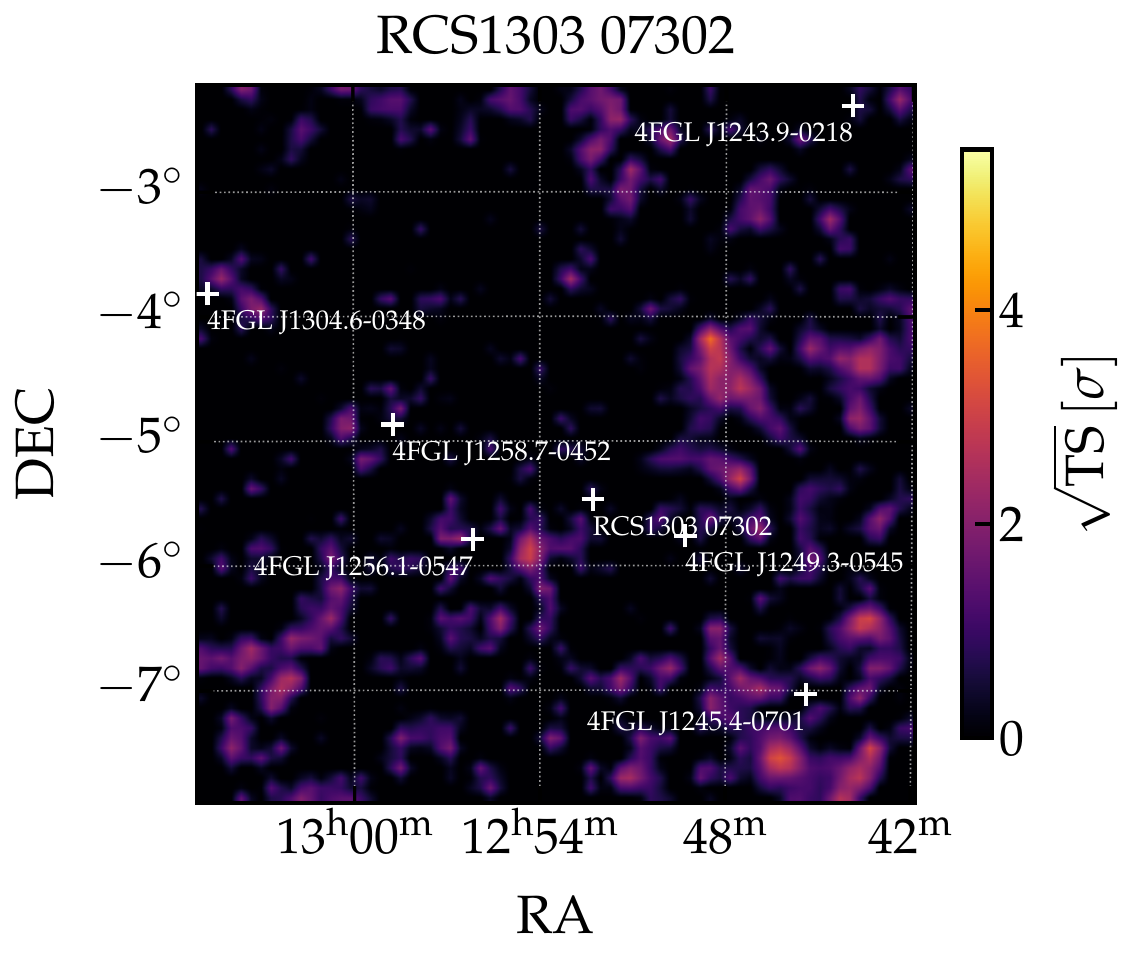}
    \end{subfigure}
\end{figure*}

\begin{figure*}[!htbp]
    \ContinuedFloat
    \centering    
    \begin{subfigure}{0.48\textwidth}
        \includegraphics[width=\linewidth]{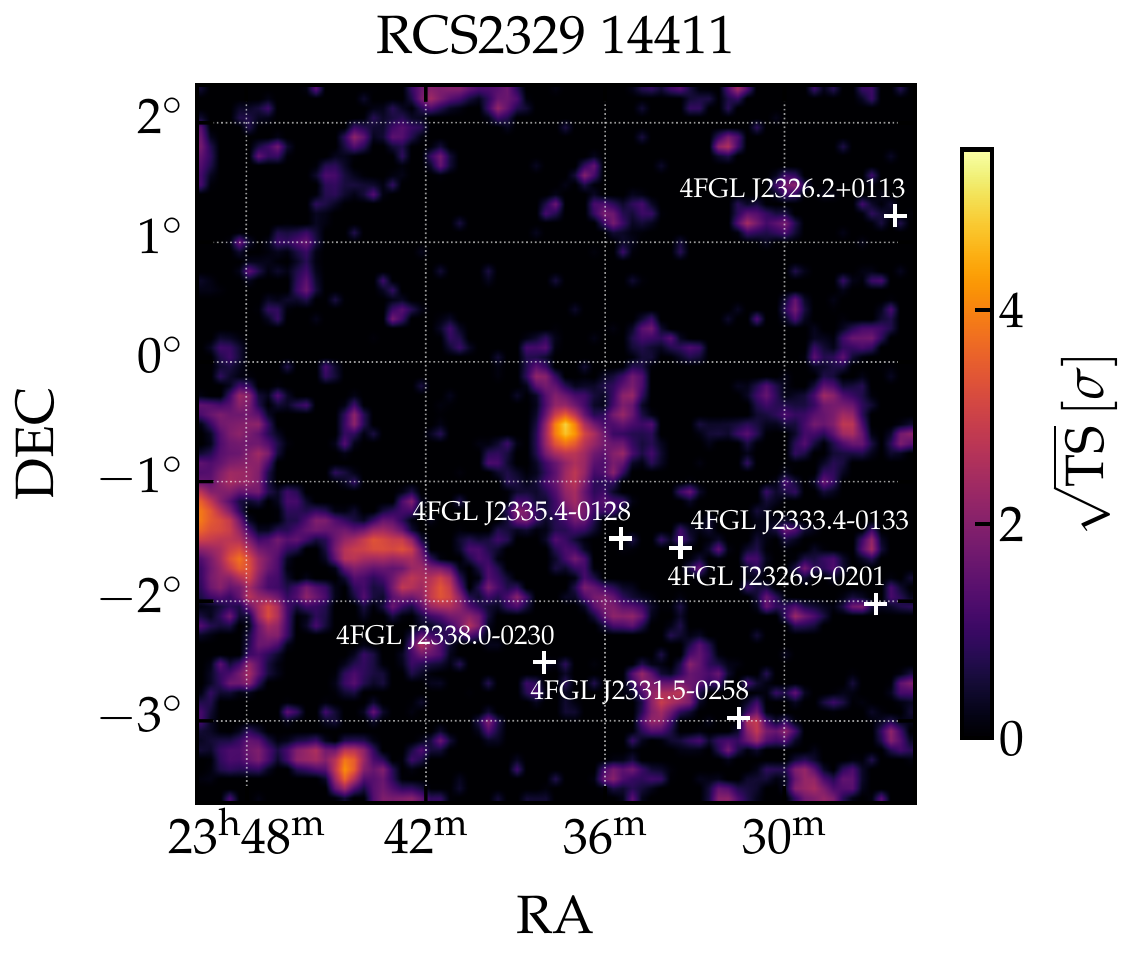}
    \end{subfigure}
    \hfill
    \begin{subfigure}{0.48\textwidth}
        \includegraphics[width=\linewidth]{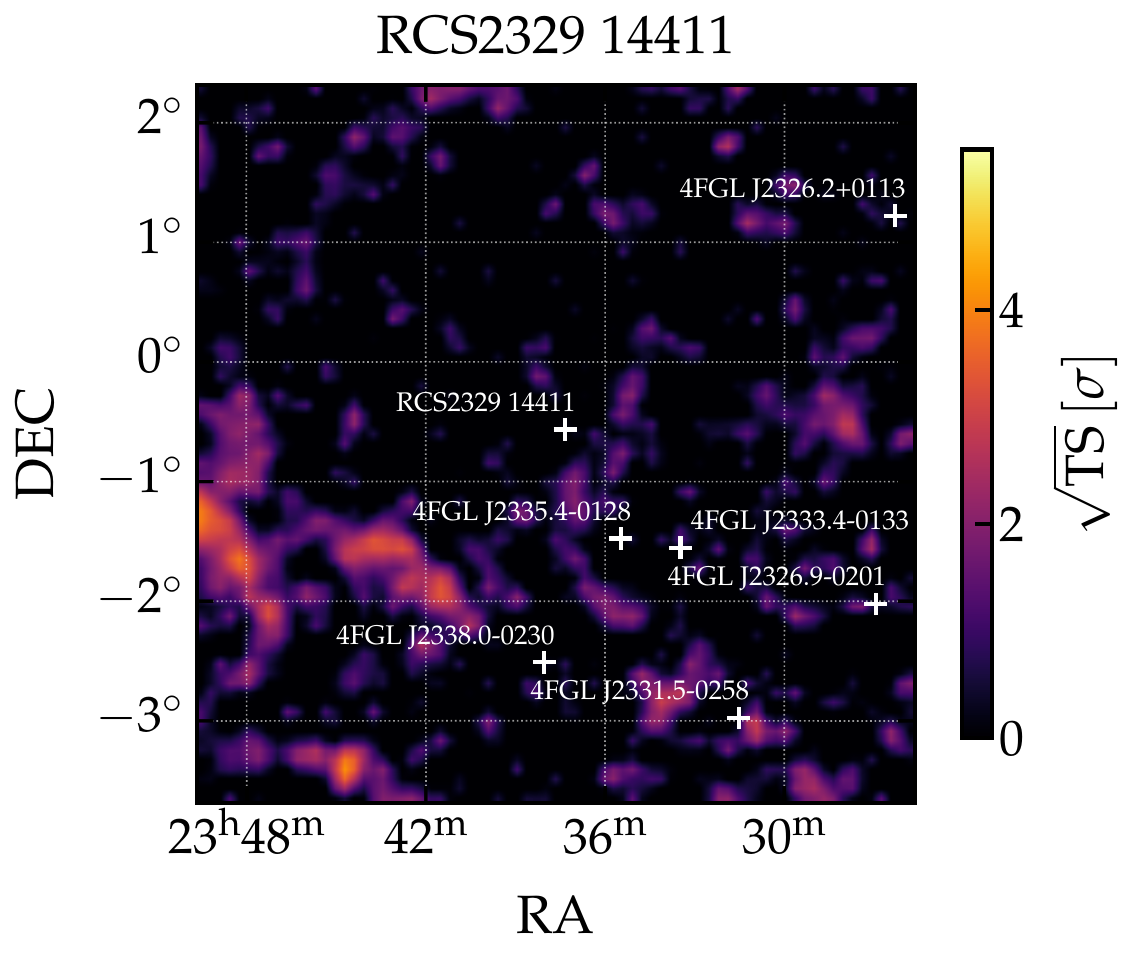}
    \end{subfigure}
\end{figure*}

\begin{figure*}[!htbp]
    \ContinuedFloat
    \centering    
    \begin{subfigure}{0.48\textwidth}
        \includegraphics[width=\linewidth]{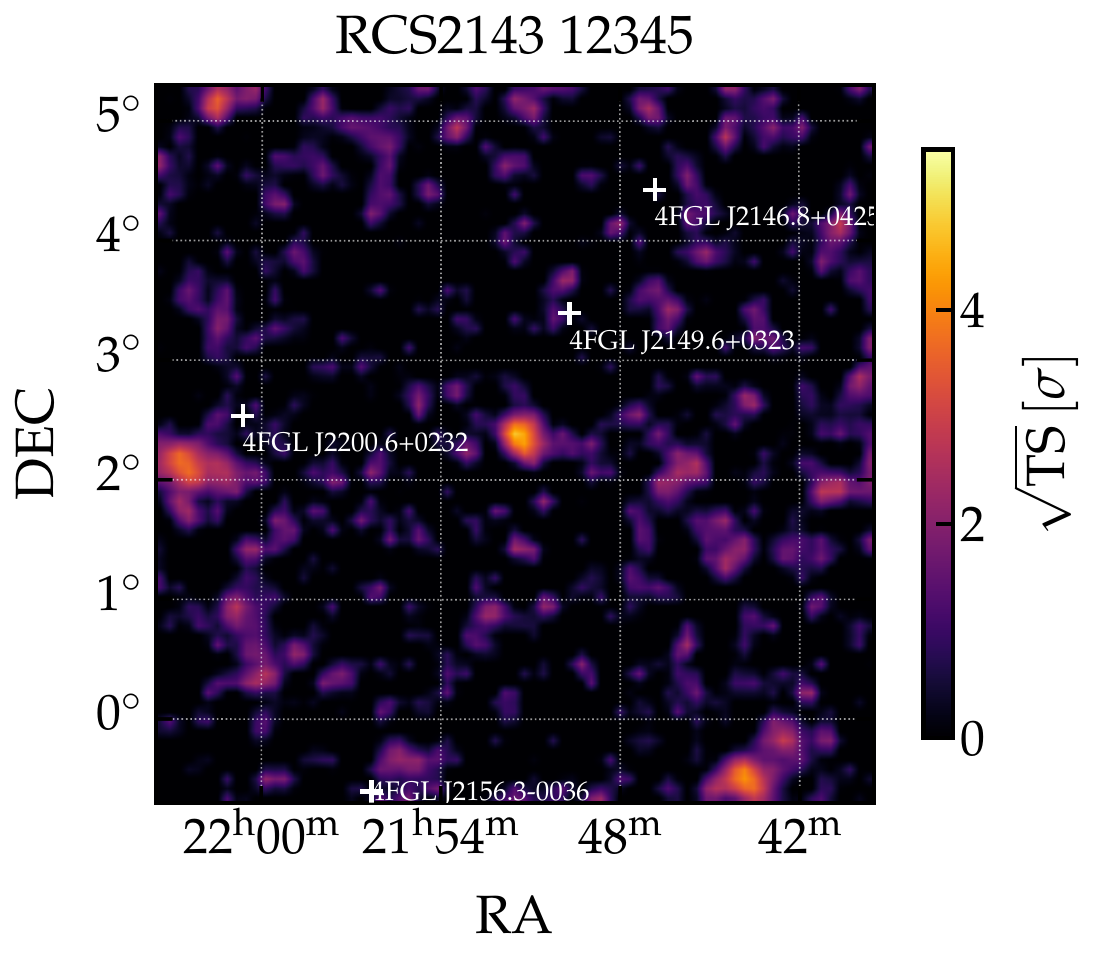}
    \end{subfigure}
    \hfill
    \begin{subfigure}{0.48\textwidth}
        \includegraphics[width=\linewidth]{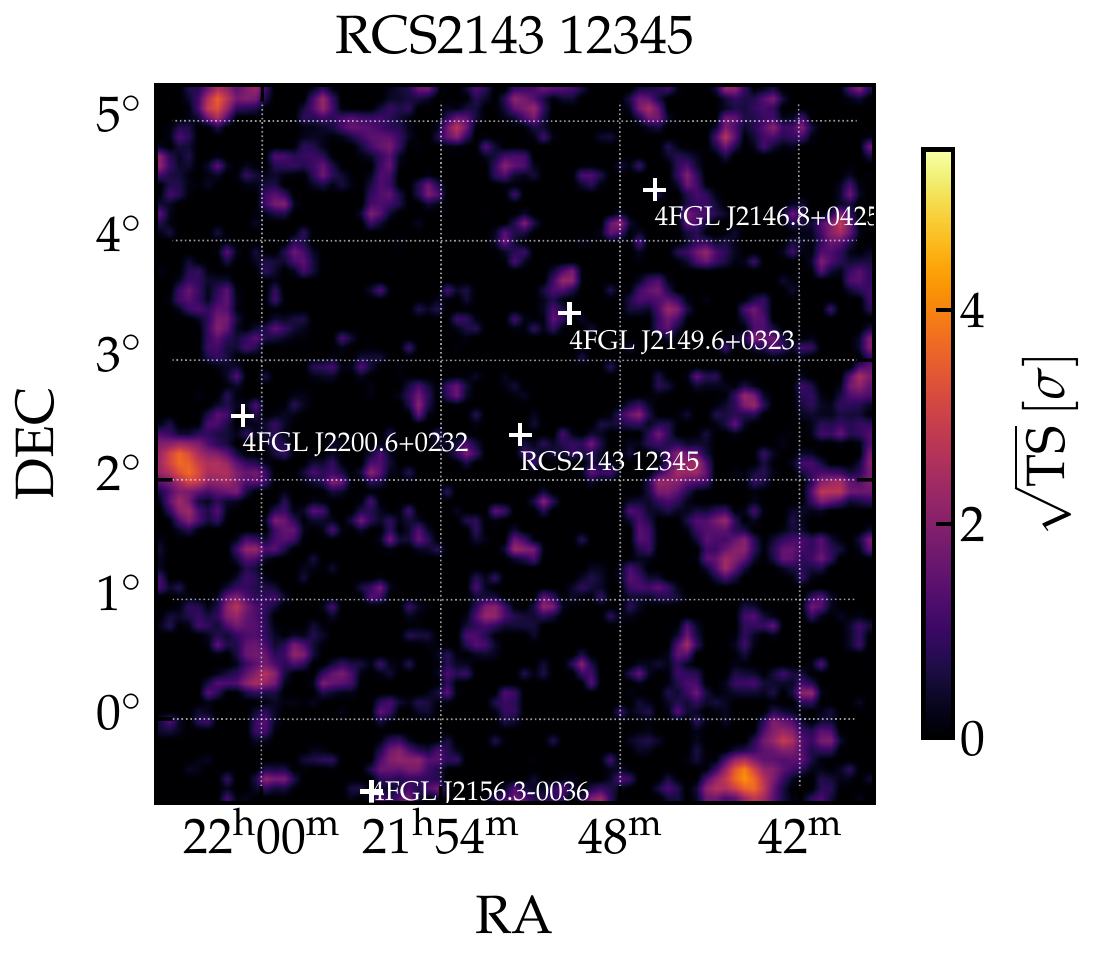}
    \end{subfigure}
\end{figure*}
\begin{figure*}[!htbp]
    \ContinuedFloat
    \centering
    \begin{subfigure}{0.48\textwidth}
        \includegraphics[width=\linewidth]{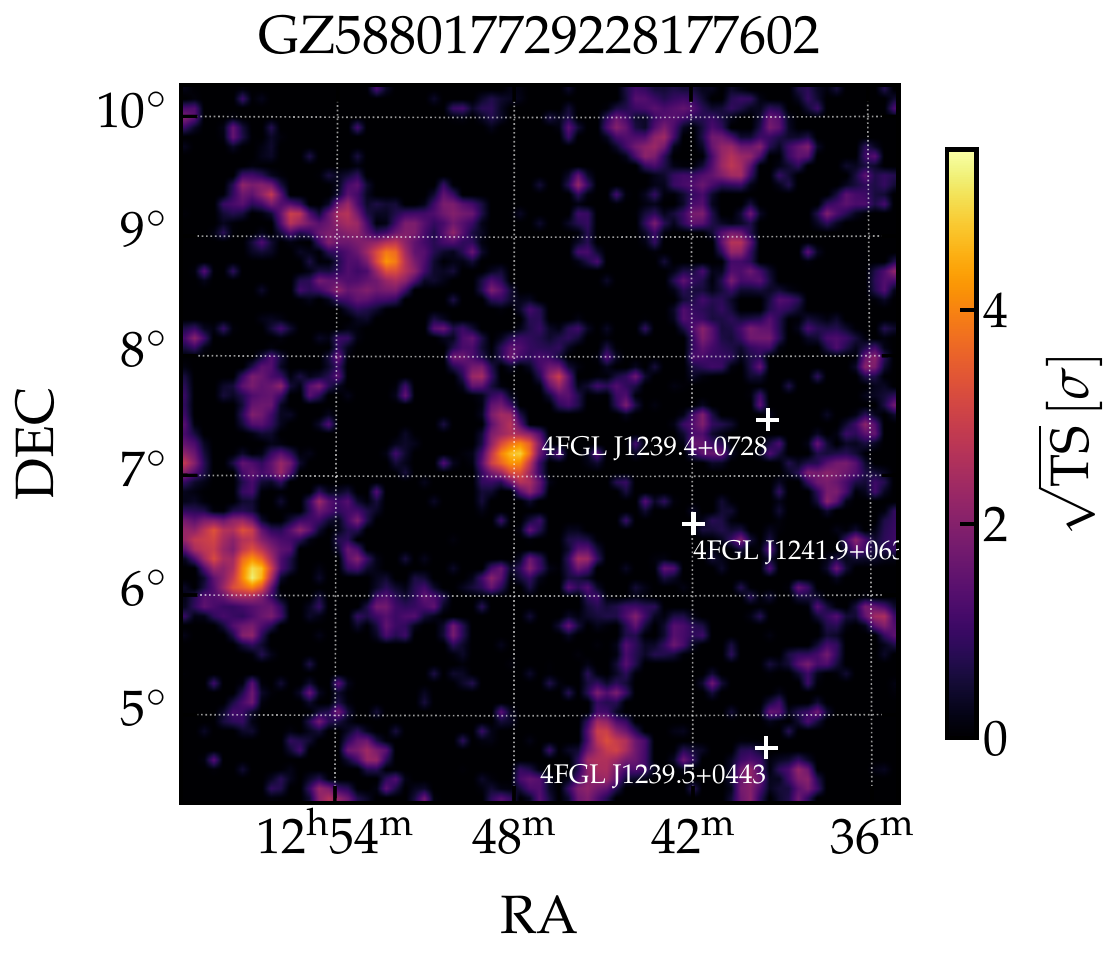}
    \end{subfigure}
    \hfill
    \begin{subfigure}{0.48\textwidth}
        \includegraphics[width=\linewidth]{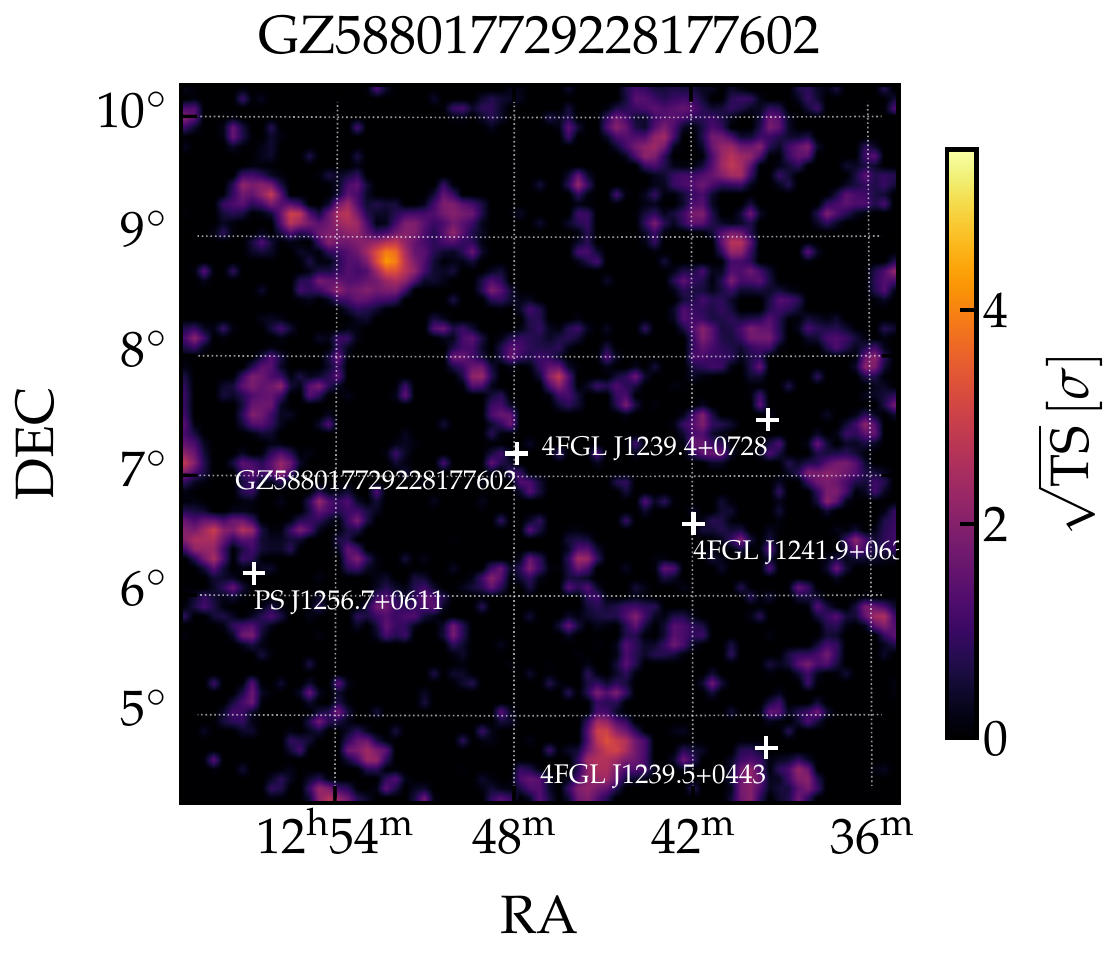}
    \end{subfigure}
    
    \caption{TS maps for galaxy mergers emitting $\gamma$ rays in the energy range 1–500 GeV with TS$>25$. For each source, the left panel shows the TS map without a test source at the center of ROI, while the right panel displays the TS map with a test source inserted in the model using a simple \texttt{PowerLaw} spectrum. The positions of the galaxy mergers in the right panels represent the localized positions. The titles of the figures correspond to the names of the sources. }
    \label{fig:tsmaps_part2}
\end{figure*}
\FloatBarrier

\clearpage
\section{SED of significant Galaxy Mergers}

\begin{figure*}[!htbp]
    \centering

    \begin{subfigure}{0.32\textwidth}
        \caption*{  \hspace{1.5cm} GT1237664877803077714}\includegraphics[width=\linewidth]{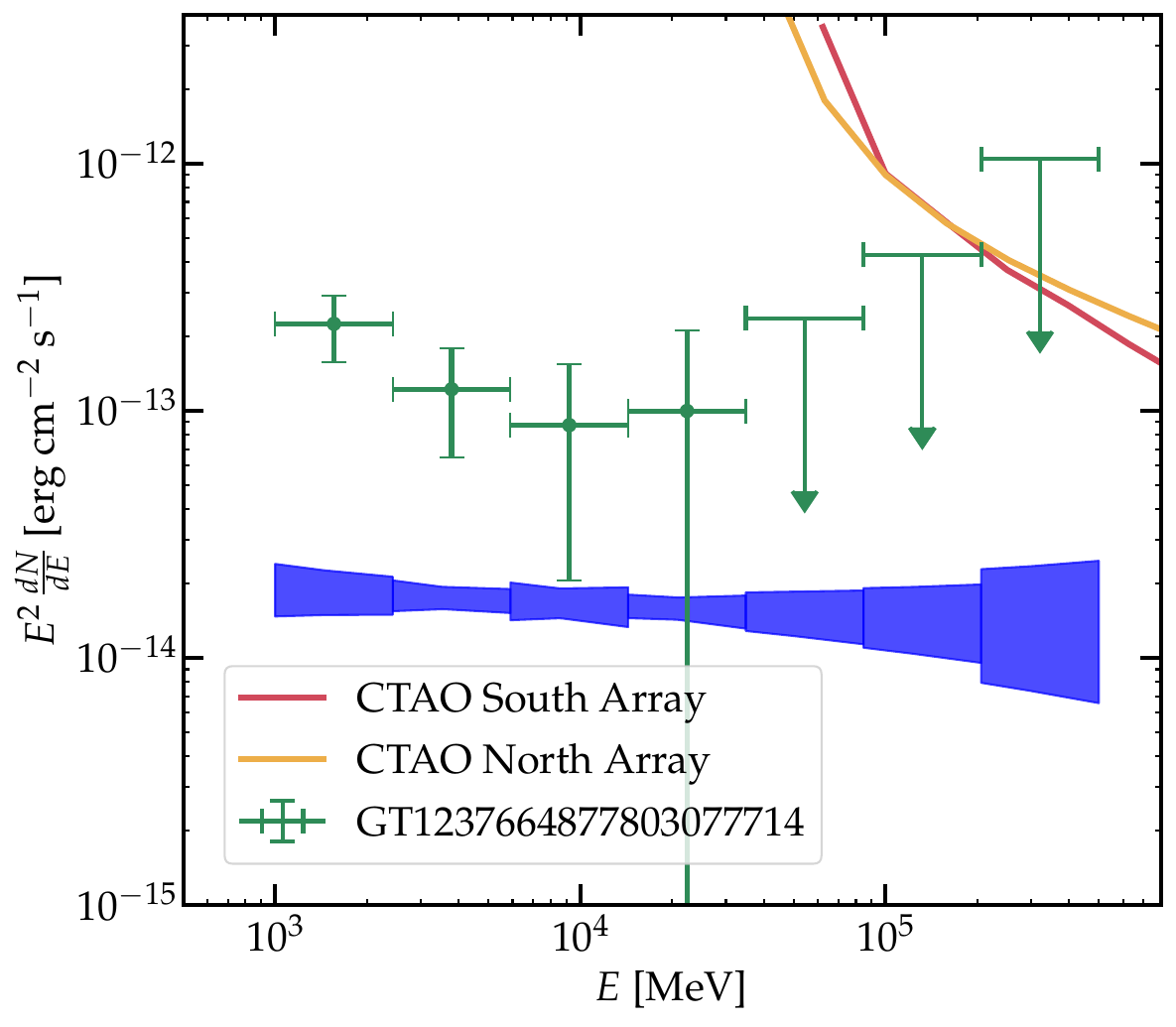}
        
    \end{subfigure}\hfill
    \begin{subfigure}{0.32\textwidth}
        \caption*{\hspace{1.7cm} GZ587742060519489665}\includegraphics[width=\linewidth]{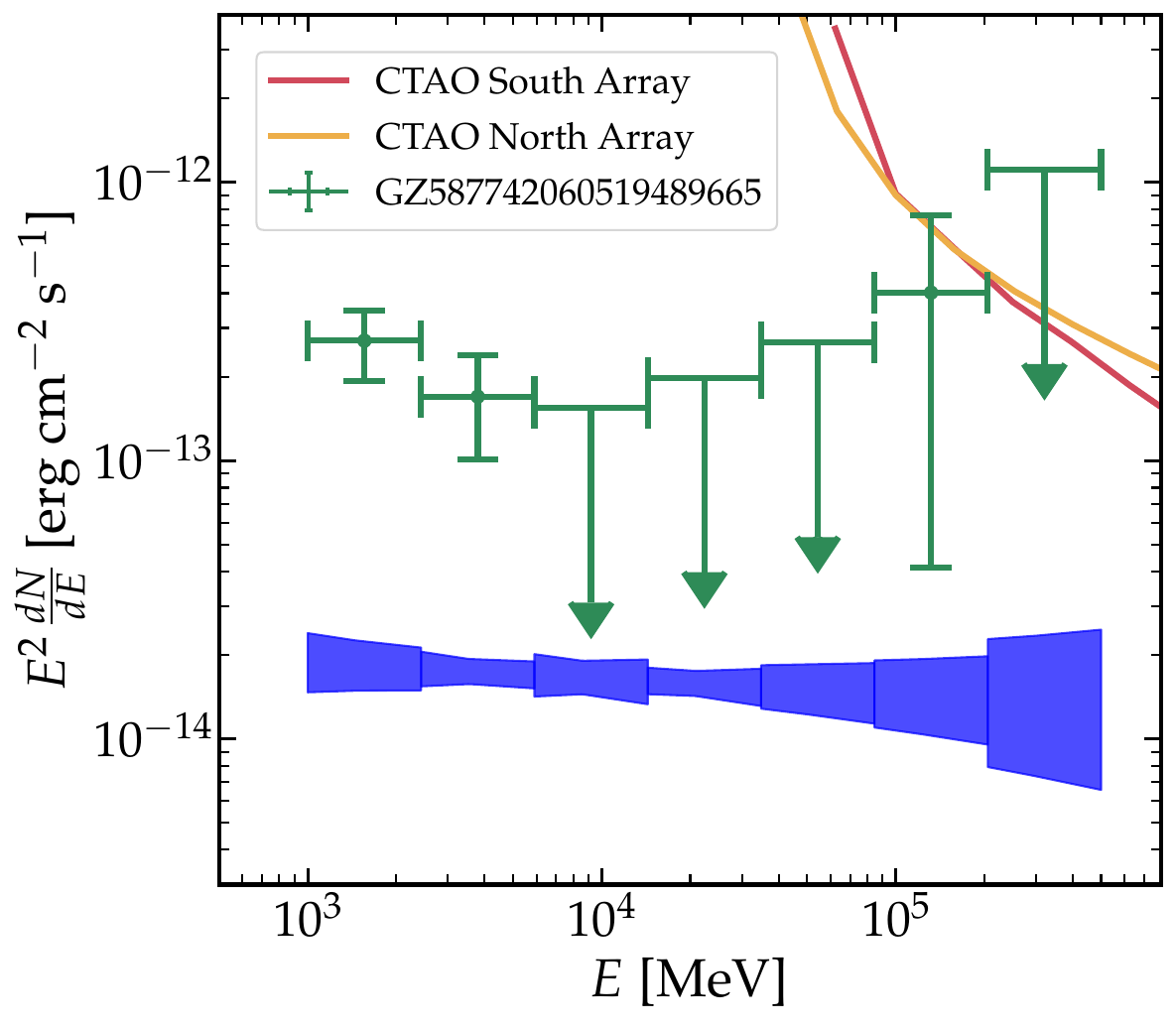}
        
    \end{subfigure}\hfill
    \begin{subfigure}{0.32\textwidth}
        \caption*{\hspace{2.2cm} RCS1645 10412}\includegraphics[width=\linewidth]{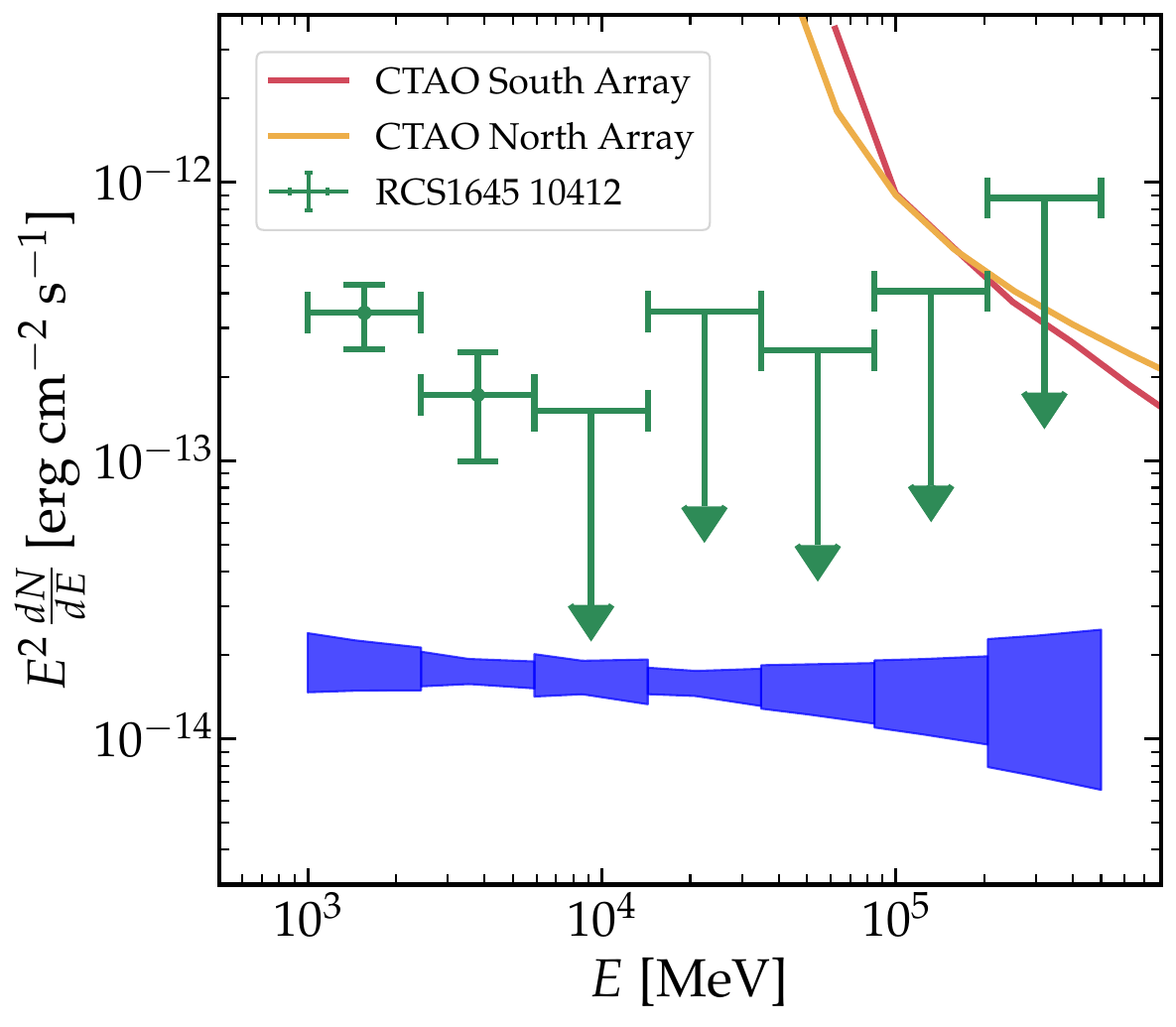}
        
    \end{subfigure}

    \begin{subfigure}{0.32\textwidth}
       \caption*{\hspace{2.2cm}RCS1303 07302}\includegraphics[width=\linewidth]{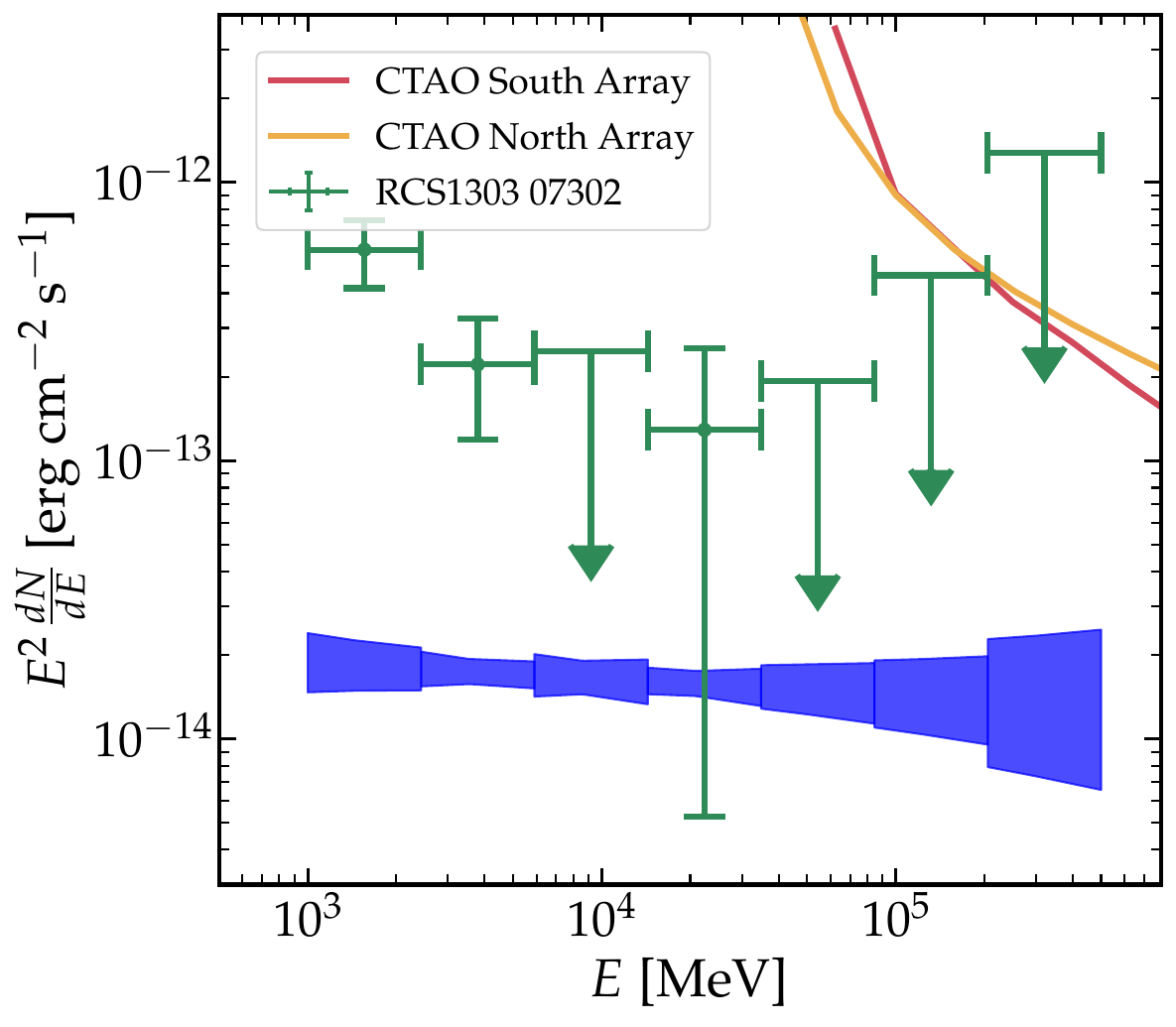}
    \end{subfigure}\hfill
    \begin{subfigure}{0.32\textwidth}
        \caption*{\hspace{2.2cm}RCS2329 14411}\includegraphics[width=\linewidth]{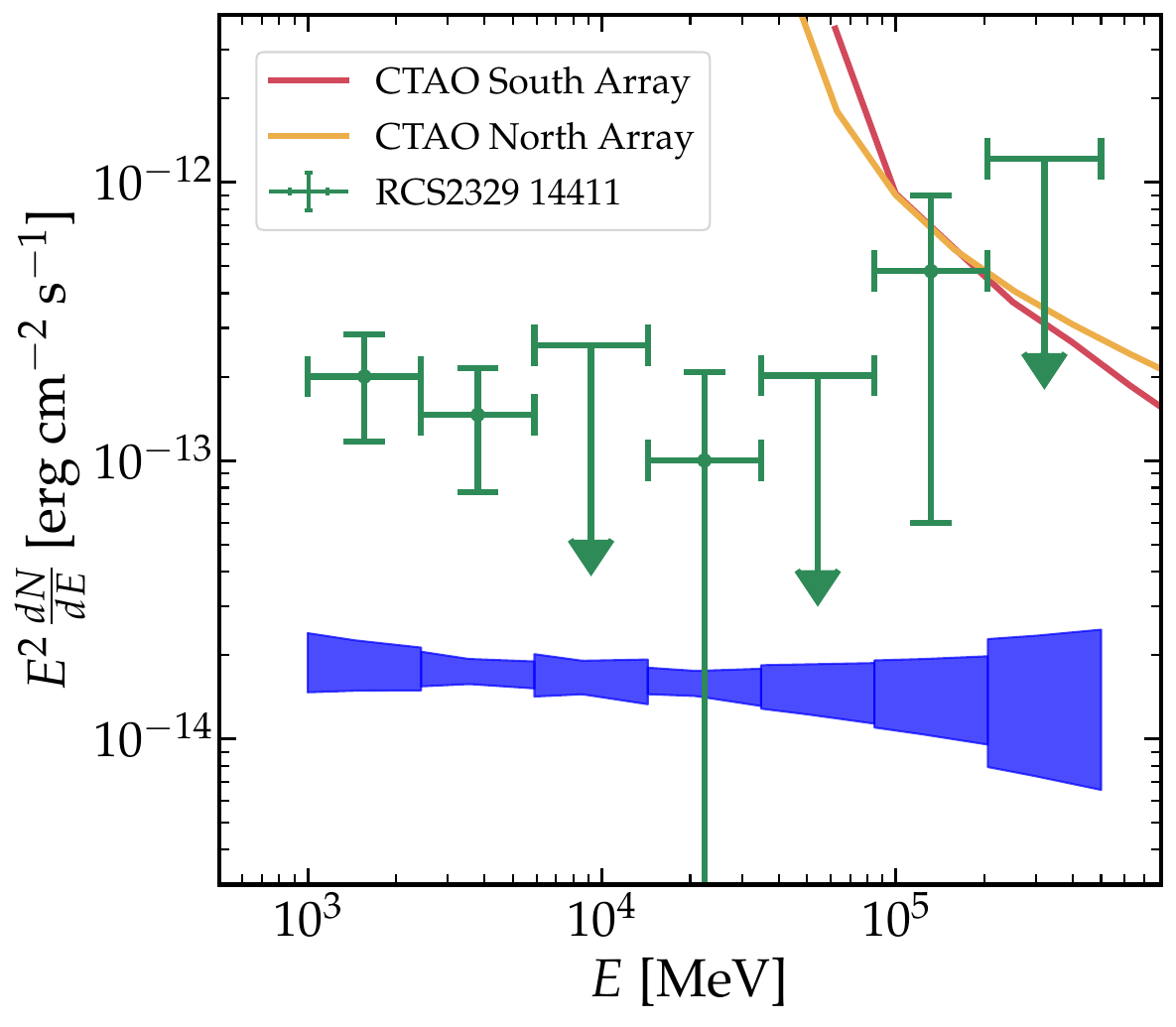}
    \end{subfigure}\hfill
    \begin{subfigure}{0.32\textwidth}
    \caption*{\hspace{2.2cm}RCS2143 12345}
        \includegraphics[width=\linewidth]{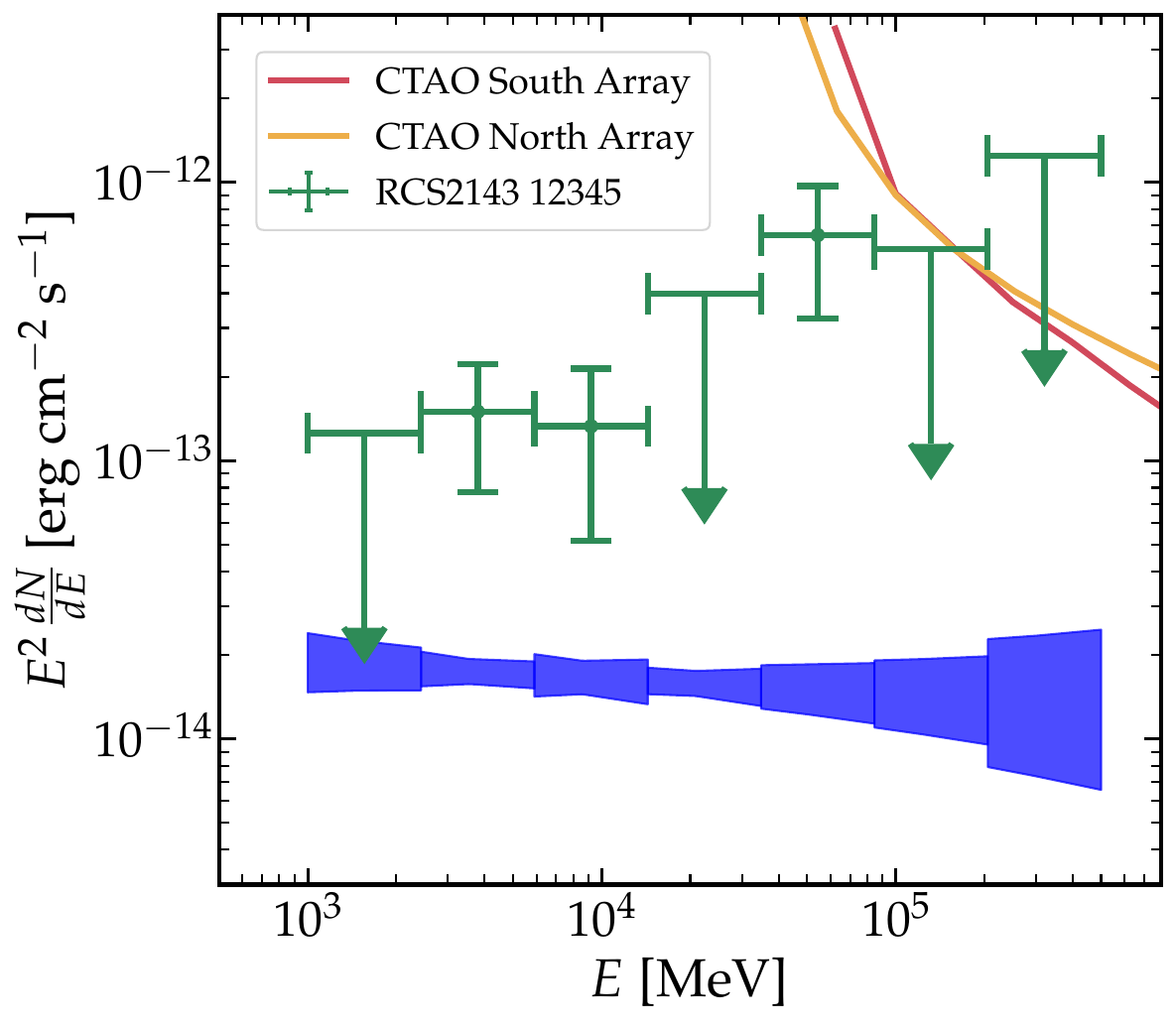}
    \end{subfigure}

    \begin{subfigure}{0.32\textwidth}
        \caption*{\hspace{1.7cm}GZ588017729228177602}
        \includegraphics[width=\linewidth]{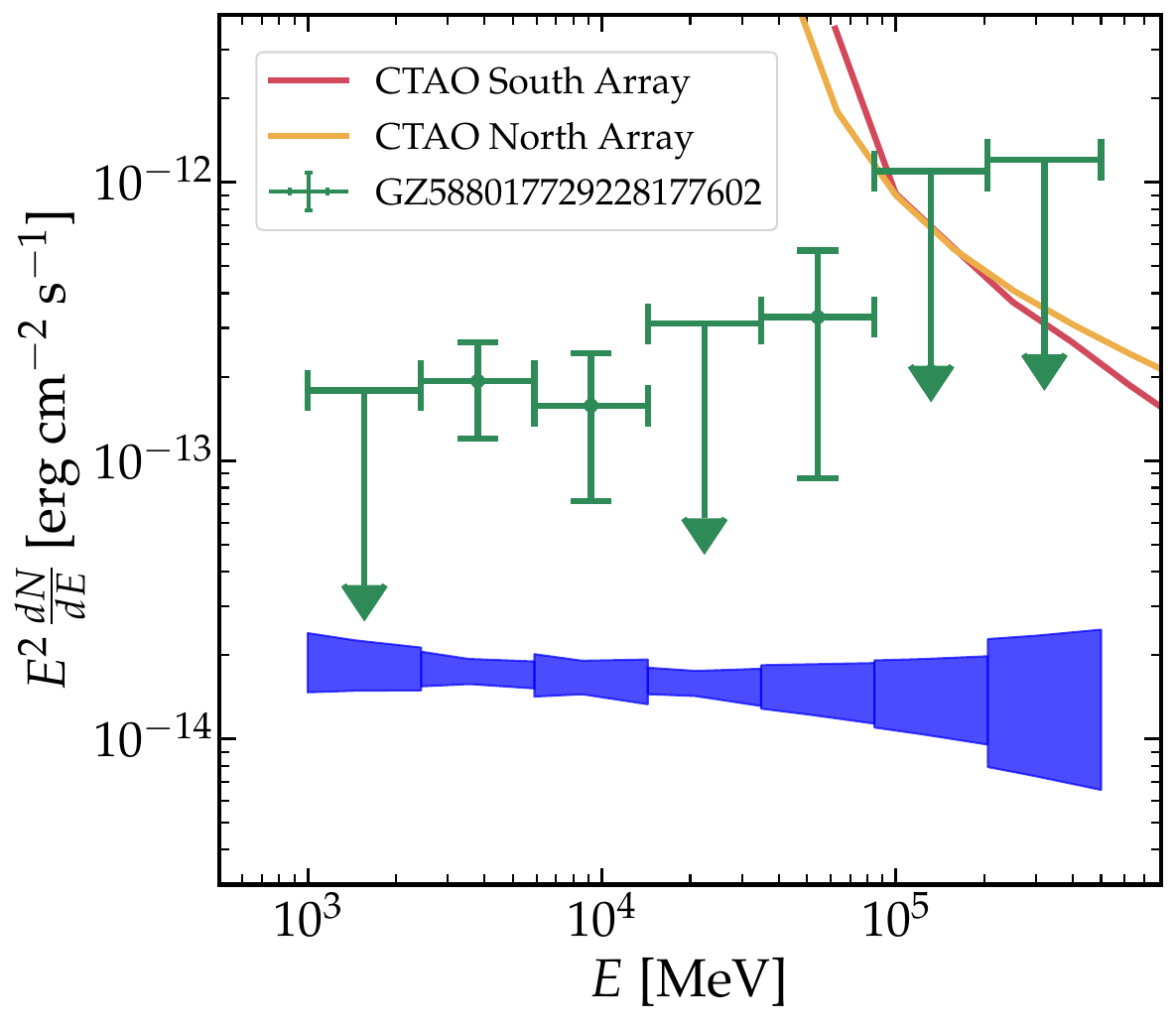}
    \end{subfigure}

    \caption{SED of the most significant sources in our analysis. The green data points show the flux along with the error bars. The upper limits on the flux are denoted by sea green arrows where the TS$<$4 for any particular energy bin. }
    \label{fig:highTS_SEDs}
\end{figure*}

\newpage
\section{List of Photon Events above 200 GeV from Galaxy Mergers}

\begin{table*}[!htbp]
\centering
\begin{threeparttable}
\sisetup{
    separate-uncertainty,
    table-align-text-post=false
}
\setlength{\tabcolsep}{7pt}
\renewcommand{\arraystretch}{1.0}

\begin{tabular}{
    l
    c
    S[table-format=3.2]
    c
    S[table-format=1.3]
    c
    c
}
\toprule
{Source Name} & {Event Coordinates} & {Event Energy} & {Event date} & {Event Prob.} & Redshift & TS\\
& {(RA, DEC) [deg]} & {[GeV]} & {[yyyy-mm-dd]} & & &\\
\midrule
RCS2143 12477$^{b}$ & (328.38, $-$4.53) & 475.39 & 2012-08-26 & 0.989 & $-$ & 6.88\\
GZ587732772132880500$^{c}$ & (190.67, 10.38) & 471.09 & 2009-07-20 & 0.998 & 0.097 & 14.84\\
RCS0047 00670$^{b}$ & (10.81, 0.08) & 437.30 & 2020-08-01 & 0.957 & $-$ & 2.89\\
GZ587741724970909786$^{c}$ & (191.91, 23.90) & 418.27 & 2008-08-25 & 0.976 & 0.083 & 6.50\\
UN587732483815375001$^{a}$ & (157.76, 46.43) & 378.42 & 2011-09-19 & 0.996 & 0.184 & 12.25\\
RCS1613 09492$^{b}$ & (245.48, 52.83) & 371.64 & 2013-05-20 & 0.966 & $-$ & 11.56\\
UN587739116320063602$^{a}$ & (144.77, 31.58) & 369.01 & 2015-11-01 & 0.953 & 0.112 & 1.96\\
29B 0920-050258$^{d}$ & (140.86, 37.38) & 348.21 & 2011-05-21 & 0.904 & 0.136 & 7.55\\
GZ587736809916858476$^{c}$ & (218.79, 13.67) & 348.41 & 2019-10-11 & 0.955 & 0.098 & 3.25\\
GZ587739131341963479$^{c}$ & (205.85, 35.09) & 335.45 & 2015-08-13 & 0.992 & 0.031 & 7.48\\
GZ587732470386720782$^{c}$ & (127.88, 32.33) & 292.09 & 2017-12-26 & 0.997 & 0.057 & 11.63\\
GZ587741722826047526$^{c}$ & (200.84, 27.15) & 285.75 & 2024-11-15 & 0.946 & 0.099 & 3.91\\
RCS2338 14329$^{b}$ & (354.14, $-$9.52) & 279.89 & 2013-02-17 & 0.977 & $-$ & 8.45\\
RCS2143 12807$^{b}$ & (329.70, $-$1.45) & 257.40 & 2012-03-13 & 0.927 & $-$ & 6.00\\
RCS1303 07417$^{b}$ & (193.97, 2.72) & 257.57 & 2008-10-18 & 0.980 & $-$ & 12.28\\
GZ587739407878652161$^{c}$ & (242.47, 21.22) & 255.70 & 2013-07-31 & 0.952 & 0.091 & 6.50\\
GZ587732471463084163$^{c}$ & (132.59, 36.43) & 260.12 & 2018-11-04 & 0.996 & 0.022 & 0.62\\
RCS2143 10925$^{b}$ & (322.39, 0.550) & 239.04 & 2022-06-02 & 0.988 & $-$ & 18.87\\
RCS2329 13748$^{b}$ & (352.17, $-$0.27) & 237.27 & 2012-08-07 & 0.925 & $-$ & 0\\
UN587725469596778656$^{a}$ & (133.29, 52.67) & 236.37 & 2011-10-15 & 0.995 & 0.063 & 8.65\\
GZ587725590382903536$^{c}$ & (261.33, 62.320) & 232.17 & 2010-07-01 & 0.981 & 0.027 & 11.69\\
GZ587733609093595484$^{c}$ & (252.94, 33.92) & 227.19 & 2022-02-08 & 0.901 & 0.084 & 12.72\\
RCS0357 05456$^{b}$ & (59.41, $-$10.82) & 229.75 & 2008-10-26 & 0.980 & $-$ & 4.98\\
RCS2338 14019$^{b}$ & (352.68, $-$13.32) & 204.18 & 2010-07-19 & 0.994 & $-$ & 0.23\\
GZ587731887344910666$^{c}$ & (120.15, 34.55) & 203.22 & 2011-12-21 & 0.933 & 0.081 & 3.66\\
RCS2329 14359$^{b}$ & (354.20, 1.58) & 209.32 & 2010-02-11 & 0.956 & $-$ & 5.18\\
RCS0310 03914$^{b}$ & (45.53, $-$14.46) & 201.63 & 2023-07-27 & 0.941 & $-$ & 7.33\\
RCS0047 00944$^{b}$ & (11.86, 1.74) & 201.13 & 2013-06-09 & 0.972 & $-$ & 3.96\\
\bottomrule
\end{tabular}
\begin{tablenotes}
\footnotesize
\item $^{a}$~\cite{2024MNRAS.533.2547F}; $^{b}$~\cite{2009yCat..21810233H}; $^{c}$~\cite{2018MNRAS.479..415A}; $^{d}$~\cite{Patton:2005qd}
\end{tablenotes}
\end{threeparttable}
\caption{High-energy photon events from 1,206 faint galaxy mergers detected above 200\,GeV, each with a detection probability greater than 90\%. A total of 28 events are listed with their celestial coordinates, event energies (in descending order), detection dates, associated event probability, redshift and TS of the sources. The precise redshift of RCS galaxy mergers is unknown~\cite{2009yCat..21810233H}. Note that most of these HE events are associated with very low TS($<4$) sources, which are consistent with statistical fluctuations of HE diffuse $\gamma$-ray photons.}
\label{tab:200GeV}
\end{table*}

\newpage
\section{List Galaxy Mergers Associated with 4FGL-DR4 Unassociated Sources}

\begin{table*}[!htbp]
\centering
\begin{threeparttable}
{\fontsize{10}{10}\selectfont
\setlength{\tabcolsep}{8pt}
\renewcommand{\arraystretch}{1.3} 
\begin{tabular}{lccccc}
\toprule
\text{4FGL-DR4 Source} & \text{RA, Dec} & \text{95\% Conf. Semimajor} & \text{Galaxy Merger} & \text{Separation}\\
 & [deg] & [deg] & & [deg]\\
\midrule
J1510.9+0551 & (227.739, 5.856) & 0.061 & RCS1514 08219$^{a}$ & 0.050\\[4pt]
J2122.5+0345 & (320.634, 3.759) & 0.223 & RCS2143 10523$^{a}$ & 0.215\\
& & & RCS2143 10558$^{a}$ & 0.109\\
& & & RCS2143 10566$^{a}$ & 0.160\\
& & & RCS2143 10589$^{a}$ & 0.194\\[4pt]
J2134.3-0319 & (323.575, $-$3.323) & 0.150 & RCS2143 11174$^{a}$ & 0.144\\[4pt]
J1009.5-0034 & (152.383, $-$0.578) & 0.141 & J100917.04-003053.01$^{b}$ & 0.089\\[4pt]
J1730.5+6018 & (262.636, 60.301) & 0.500 & J173127.36+604133.72$^{b}$ & 0.407\\
& & & J172638.64+601357.00$^{b}$ & 0.486\\
& & & J172647.52+602623.64$^{b}$ & 0.483\\
& & & J173032.64+600101.56$^{b}$ & 0.284\\
& & & UN587725577498722606$^{c}$ & 0.472\\
& & & UN587725577499115707$^{c}$ & 0.403\\
& & & UN587725590920823056$^{c}$ & 0.370\\
& & & UN587725590920495410$^{c}$ & 0.388\\[4pt]
J1014.6+6126 & (153.658, 61.437) & 0.099 & UN587725474418262141$^{c}$ & 0.087\\[4pt]
J0743.2+4034 & (115.800, 40.578) & 0.253 & UN587725551729639787$^{c}$ & 0.229\\
& & & UN587725551729574203$^{c}$ &  0.229\\
& & & UN587725774530347146$^{c}$ &  0.099\\[4pt]
J0838.5+4013 & (129.629, 40.22) & 0.158 & UN587731522815197379$^{c}$ &  0.105\\
& & & UN587731872316784916$^{c}$ & 0.150\\[4pt]
J1122.2+3926 & (170.56, 39.45) & 0.199 & UN587738947752689844$^{c}$ & 0.096\\[4pt]
J1007.6+4253 & (151.902, 42.889) & 0.282 & UN588297864722448543$^{c}$ & 0.246\\
& & & UN588297864722317376$^{c}$ & 0.046\\
& & & UN588297864722317417$^{c}$ & 0.199\\[4pt]
J1727.1+5955 & (261.776, 59.925) & 0.042 & UN587725576962179298$^{c}$ & 0.037\\[4pt]
J1417.3+6059 & (214.342, 60.992) & 0.102 & UN587728917910520039$^{c}$ & 0.095\\[4pt]
J0800.5+3715 & (120.143, 37.259) & 0.098 & UN587728906638393631$^{c}$ & 0.081\\[4pt]
J1346.5+5330 & (206.648, 53.502) & 0.116 & UN587733411517628523$^{c}$ & 0.104\\[4pt]
J0918.5+3516 & (139.645, 35.273) & 0.16 & UN587735043617128456$^{c}$ & 0.133\\[4pt]
J1359.9+3616 & (209.991, 36.283) & 0.181 & UN587736940369346661$^{c}$ & 0.176\\[4pt]
J1424.8+4910 & (216.202, 49.179) & 0.435 & UN588017713124409476$^{c}$ & 0.401\\
& & & UN588017713661149414$^{c}$ & 0.257\\[4pt]
J1428.6+0705 & (217.169, 7.090) & 0.366 & GZ587730023864205514$^{d}$ & 0.179\\
\bottomrule
\end{tabular}}
\begin{tablenotes}
\footnotesize
\item $^{a}$~\cite{2009yCat..21810233H}; $^{b}$~\cite{Allam:2003me}; $^{c}$~\cite{2024MNRAS.533.2547F}; $^{d}$~\cite{2018MNRAS.479..415A}
\end{tablenotes}
\vspace{0.5em}
\caption{List of 18 unassociated 4FGL–DR4 sources that show potential association with 34 galaxy mergers from our combined catalog of 31,464 galaxy mergers. Associated galaxy mergers lie within the 95\% confidence semimajor axis of their respective unassociated sources.}
\label{tab:unassocia}
\end{threeparttable}
\end{table*}
\end{document}